\documentclass[a4paper,11pt]{article}
\usepackage{jcappub} % for details on the use of the package, please see the JINST-author-manual
\usepackage{graphicx}
\usepackage[varg]{txfonts}
\usepackage{graphicx}
\usepackage{psfrag}
\usepackage{hyperref}
\usepackage{gensymb}
\usepackage{xcolor}
\usepackage{tabularx}
\usepackage{lipsum}
\usepackage[normalem]{ulem}
\usepackage{amsmath}
\usepackage[labelfont=bf, labelsep=period]{caption}
\usepackage{subcaption}
\usepackage{mathtools}
\usepackage{balance}
\usepackage{adjustbox}
\usepackage{float} 
\usepackage{placeins}
\usepackage{caption}
\usepackage{stfloats}
\usepackage{pifont}
\usepackage{xcolor}
\usepackage{dsfont}
\usepackage{multirow}
\usepackage{changepage}
\usepackage{makecell}
\usepackage{soul}

\usepackage[colorinlistoftodos]{todonotes}
\usepackage{easyReview}

\usepackage{enumitem}

\usepackage{siunitx} 
\usepackage[switch,displaymath,mathlines]{lineno}

\newcommand{\lb}{\emph{LiteBIRD}}
\title{\boldmath Debiasing cosmological parameters from large-scale foreground contamination in Cosmic Microwave Background data}

\author{Alessandro Carones}
\affiliation{\it International School for Advanced Studies (SISSA), Via Bonomea 265, Trieste 34136, Italy}
\affiliation{\it INFN Sezione di Trieste, 
Via Valerio 2, Trieste 34127, Italy}
\affiliation{\it IFPU,
Via Beirut 2, Trieste 34151, Italy}

\emailAdd{acarones@sissa.it}

\abstract{Current and future Cosmic Microwave Background (CMB) experiments aim to achieve high-precision reconstruction of the CMB polarization signal, with the most ambitious objective being the detection of primordial $B$ modes sourced by cosmic inflation. Given the expected low amplitude of the signal, its estimate—parametrized by the tensor-to-scalar ratio $r$—is highly susceptible to contamination from Galactic foreground residuals that remain after component separation.  

In this work, we introduce an agnostic, model-independent procedure to construct a spectral template of residual foreground contamination in the observed angular power spectrum. Specifically, a cleaned multifrequency set of foreground-emission maps is blindly reconstructed from the observed data using the Generalized Needlet Internal Linear Combination (GNILC) technique. These maps are then combined with the weights adopted for CMB reconstruction, yielding an estimate of the spatial distribution of foreground residuals after component separation. The power spectrum of this estimated residual map, after proper noise debiasing, is incorporated into the spectral model of the cosmological likelihood, thereby enabling unbiased inference of cosmological parameters.

We validate the proposed method using realistic simulations of a \emph{LiteBIRD}-like experiment processed with two alternative Internal Linear Combination (ILC) component-separation techniques, focusing on constraints on the tensor-to-scalar ratio. When the foreground contribution is not included in the spectral model, the resulting posteriors of the cosmological parameter are biased, irrespective of its input value, the assumed foreground model, or the adopted masking strategy. Conversely, when the residual template is included in the likelihood, the analysis yields unbiased estimates of $r$ for all considered cases, thereby demonstrating the robustness of the proposed procedure.

The pipeline has been made publicly available as part of the \href{https://github.com/alecarones/broom}{\textcolor{black}{BROOM}} Python package.
}

\keywords{CMBR polarization -- gravitational waves and CMBR polarization -- cosmological parameters from CMBR -- CMBR experiments}

\begin{document}
\maketitle
\flushbottom

\section{Introduction}
\label{sec:intro}

Upcoming measurements of the polarization field of the Cosmic Microwave Background (CMB), with unprecedented sensitivity, promise to deepen our understanding of the Universe \cite{SO_2019,PTEP}. For instance, they will provide new insights into cosmic inflation \cite{2015APh....63...55A,2016ARA&A..54..227K}, the reionization history \cite{1997PhRvD..55.1822Z}, neutrino properties \cite{2004PhRvD..69h3002B,2006PhR...429..307L}, additional relativistic relics \cite{2016JCAP...01..007B, Planck_cosmopars}, and cosmic birefringence \cite{2009PhRvL.102k1302K,2020PhRvL.125v1301M}. These advances rely on precise reconstruction of both the CMB $E$-mode and $B$-mode polarization patterns across the sky \cite{1997PhRvL..78.2058K,1997PhRvD..55.1830Z}. Achieving this, however, is hindered by two major challenges. First, residual contamination from polarized foreground emission inevitably remains in the data due to imperfect component separation. Second, instrumental systematics can leak into the reconstructed CMB signal. Both effects can mimic genuine cosmological signatures at the power spectrum level, thereby introducing systematic biases in the inference of cosmological parameters from CMB polarization measurements. To mitigate the impact of foreground residuals, three complementary (though not mutually exclusive) strategies can be pursued:  
(i) improving the effectiveness and robustness of component-separation methods in the data-analysis pipeline;  
(ii) refining the masking of the most contaminated sky regions after component separation; and  
(iii) explicitly modelling the residual contamination at the spectral level and incorporating it into the likelihood.

In this work, we focus on the last approach. We present a new methodology for constructing a spectral template of foreground residuals after component separation, which is then consistently incorporated into the spectral likelihood to mitigate biases in cosmological parameter estimation. The approach relies on obtaining cleaned templates of foreground emission at all observed frequency channels using the Generalized Needlet Internal Linear Combination (GNILC) technique \cite{GNILC_intro, GNILC}. These templates are subsequently combined with the component separation weights employed to recover the CMB signal, yielding an estimate of the overall residual contamination in the reconstructed CMB maps. The angular power spectrum of this residual template is finally incorporated into the cosmological harmonic likelihood as a spectral model of the contamination in the observed power spectrum. The procedure is fully model-independent, ensuring robustness against the specific properties of polarized foregrounds that will be revealed by future surveys, and is broadly applicable across different map-based component separation pipelines. 

The approach is validated using realistic simulations of the microwave sky based on a \lb-like instrumental configuration as proposed in \cite{PTEP}, designed to achieve high-precision reconstruction of the CMB polarization field over a large fraction of the sky. As a proof of concept, we apply the method to the recovered $B$-mode power spectrum, demonstrating its ability to debias the estimation of the amplitude of primordial tensor perturbations, quantified by the tensor-to-scalar ratio $r$ \cite{2016ARA&A..54..227K}. The component-separation techniques adopted for CMB reconstruction in this work are state-of-the-art Internal Linear Combination (ILC) methods \cite{NILC, MCNILC}.

Details of the simulated data set, the employed component separation pipelines, and the inference of the tensor-to-scalar ratio, which form the basis of the analysis, are presented in Section \ref{sec:methods}. A full description of the methodology used to derive a template of foreground residuals and its incorporation into the likelihood is provided in Section \ref{sec:marginal}. The results are reported in Section \ref{sec:results}, followed by concluding remarks in Section \ref{sec:concl}.

\section{Background methodology}
\label{sec:methods}

In this section, we describe the considered sets of microwave sky simulations (Section \ref{ssec:sims}), which are subsequently processed with a realistic component separation pipeline (outlined in Section \ref{ssec:compsep}). To assess the quality of the cosmological information that can be extracted from the reconstructed CMB maps, we rely on the angular power spectrum, the standard statistical measure in this context. The procedure for computing the angular power spectra, including the adopted masking strategies, and the subsequent cosmological inference are detailed in Section \ref{ssec:spectra}.

\subsection{Microwave sky simulations}
\label{ssec:sims}
In this work, we simulate realistic observations expected from a \emph{LiteBIRD}-like satellite. This instrumental configuration is designed to observe the sky in 15 frequency channels, achieving an overall polarization sensitivity of $2.2\,\mu\mathrm{K}\text{-arcmin}$. For our analysis, we adopt the angular resolution and sensitivity specifications for each frequency channel as reported in Table~13 of \cite{PTEP}.

The simulated observations include contributions from the CMB signal, Galactic foreground emission, and instrumental noise. We generate 100 independent simulations of the microwave sky, each featuring different realizations of the CMB signal and noise. All components are simulated as \texttt{HEALPix}-pixelized maps \cite{healpix, healpy} at a spatial resolution of $N_{\text{side}} = 64$, in terms of the linear polarization Stokes parameters $Q$ and $U$, and expressed in the standard CMB temperature units of $\mu\mathrm{K}_{\text{CMB}}$. The choice of this \texttt{HEALPix} resolution is motivated by the focus on large angular scales. Indeed, in this proof-of-concept study, we apply the proposed debiasing technique to the estimation of the tensor-to-scalar ratio $r$, whose constraining power lies primarily at multipoles $\ell < 180$, which can be fully sampled with the chosen $N_{\text{side}}$.
 In addition, the most significant contamination from Galactic diffuse emission arises precisely on these angular scales, making them the regime where cosmological parameter inference is most susceptible to astrophysical systematic biases.

The instrumental beams are assumed to be Gaussian and the bandpasses are modeled as $\delta$ functions (i.e., with no bandwidth). These simplifying assumptions do not affect either the core methodology presented in this work or its main results.

The CMB polarization anisotropies are modeled as Gaussian random fields drawn from theoretical angular power spectra computed with the \texttt{CAMB} Boltzmann code\footnote{\url{https://github.com/cmbant/camb}} \cite{camb}, assuming the $\Lambda$CDM best-fit cosmological parameters from the Planck Data Release 3 angular power spectra \cite{Planck_cosmopars}. For this analysis, we consider three sets of CMB maps that differ only in their $B$-mode component: one with a $B$-mode angular power spectrum corresponding to lensed $E$-modes only ($r = 0$) and two with additional tensor perturbations at the levels of $r=0.004$ and $r=0.01$, respectively. These input tensor-to-scalar ratios are chosen to represent scientifically and observationally relevant benchmarks for future CMB data analyses \cite{PTEP, SO_2019, STAROBINSKY1980}.

Simulated Galactic foreground emission includes synchrotron \citep{krach_2018,2022ApJ...936...24W} and thermal dust radiation \cite{Planck2018_compsep,2015JCAP...12..020C}, generated using the \texttt{PySM} Python package \cite{pysm, pysm3, pysm2025}. Synchrotron radiation, $X_{\textrm{s}}(\hat{n},\nu)$, along the line-of-sight direction $\hat{n}$ and at frequency $\nu$, is modeled as the dominant low-frequency emission and assumed to follow a power-law spectral energy distribution (SED) in brightness temperature units:

\begin{equation}
X_{\textrm{s}}(\hat{n},\nu) = X_{\textrm{s}}(\hat{n},\nu_{\textrm{s}})\cdot
\left(\frac{\nu}{\nu_{\textrm{s}}} \right)^{\beta_{\textrm{s}}(\hat{n})},
\label{eq:sync}
\end{equation}
where $X=\{Q,U \}$ are Stokes parameters, $\beta_{\textrm{s}}(\hat{n})$ is the synchrotron spectral index and $X_{\textrm{s}}(\hat{n},\nu_{\textrm{s}})$ is the synchrotron template at the reference frequency $\nu_{\textrm{s}}$.
We consider two \texttt{PySM} synchrotron models: \texttt{s1} and \texttt{s5}. Both adopt as polarization template the \emph{WMAP} 9-year 23-GHz Q and U maps \citep{sync_temp}, smoothed with a Gaussian kernel with Full Width at Half Maximum (FWHM) equal to $5^\circ$. In \texttt{s1}, small-scale fluctuations are added as Gaussian realizations drawn from the extrapolated large-scale angular power spectrum \cite{pysm}. In contrast, \texttt{s5} introduces non-Gaussian small-scale modes using the \texttt{logpoltens} formalism \cite{pysm2025}. In both cases, the synchrotron spectral index map is obtained from a spectral fit to the Haslam 408-MHz and \emph{WMAP} 23-GHz 7-year data \citep{sync_index}, but in \texttt{s5} it is further rescaled to account for the increased variability observed in S-PASS data \cite{krach_2018}. Overall, the \texttt{s5} model represents a more complex and realistic description of Galactic synchrotron emission.

Thermal dust emission, $X_{\textrm{d}}(\hat{n},\nu)$, is modeled as a modified blackbody (MBB) spectrum in intensity units:
\begin{equation}
X_{\textrm{d}}(\hat{n},\nu) = X_{\textrm{d}}(\hat{n},\nu_{\textrm{d}}) \cdot
\left(\frac{\nu}{\nu_{\textrm{d}}} \right)^{\beta_{\textrm{d}}(\hat{n})}
\cdot \frac{B_{\nu}\left(T_{\textrm{d}}(\hat{n})\right)}{B_{\nu_{\textrm{d}}}\left(T_{\textrm{d}}(\hat{n})\right)},
\label{eq:dust}
\end{equation}
where $B_{\nu}(T)$ is the Planck blackbody function, $\beta_{\textrm{d}}(\hat{n})$ is the dust spectral index, $T_{\textrm{d}}(\hat{n})$ the dust temperature, and $X_{\textrm{d}}(\hat{n},\nu_{\textrm{d}})$ the dust template at a reference frequency $\nu_{\textrm{d}}$.
As for synchrotron, we consider two \texttt{PySM} dust models featuring increasing complexity: \texttt{d1} and \texttt{d10}. In both cases, the $\beta_{\textrm{d}}$, $T_{\textrm{d}}$, and $X_{\textrm{d}}(\hat{n},\nu_{\textrm{d}})$ maps are obtained from MBB fits to \emph{Planck} data. These fits are performed with the \texttt{Commander} algorithm \citep{2016A&A...594A..10P, pysm} for \texttt{d1} and using the GNILC-denoised data \citep{GNILC, pysm2025} for \texttt{d10}. Large-scale modes are directly extracted from the data, while small-scale features are added following the same prescriptions used for the \texttt{s1} and \texttt{s5} synchrotron models, respectively.

From these ingredients, we construct two distinct sets of foreground maps: a simpler one, \texttt{d1s1} (combining \texttt{d1} dust and \texttt{s1} synchrotron), and a more complex one, \texttt{d10s5} (including \texttt{d10} dust and \texttt{s5} synchrotron).

Finally, instrumental noise in the $Q$ and $U$ maps is simulated as white, isotropic, and independent Gaussian realizations, with the pixel standard deviation determined from the sensitivity specifications reported in \cite{PTEP}. The scanning strategy is not included, as it is not expected to introduce any significant impact on the results. Moreover, no correlated $1/f$ noise component is added, under the assumption of proper functioning of the Half-Wave Plate polarization modulator.

All these microwave components are generated separately through the \texttt{BROOM} Python package\footnote{\url{https://github.com/alecarones/broom}} (corresponding paper in preparation) and then co-added together.

\subsection{Component separation}
\label{ssec:compsep}
\begin{figure}
	\centering
	\includegraphics[width=0.495\textwidth]{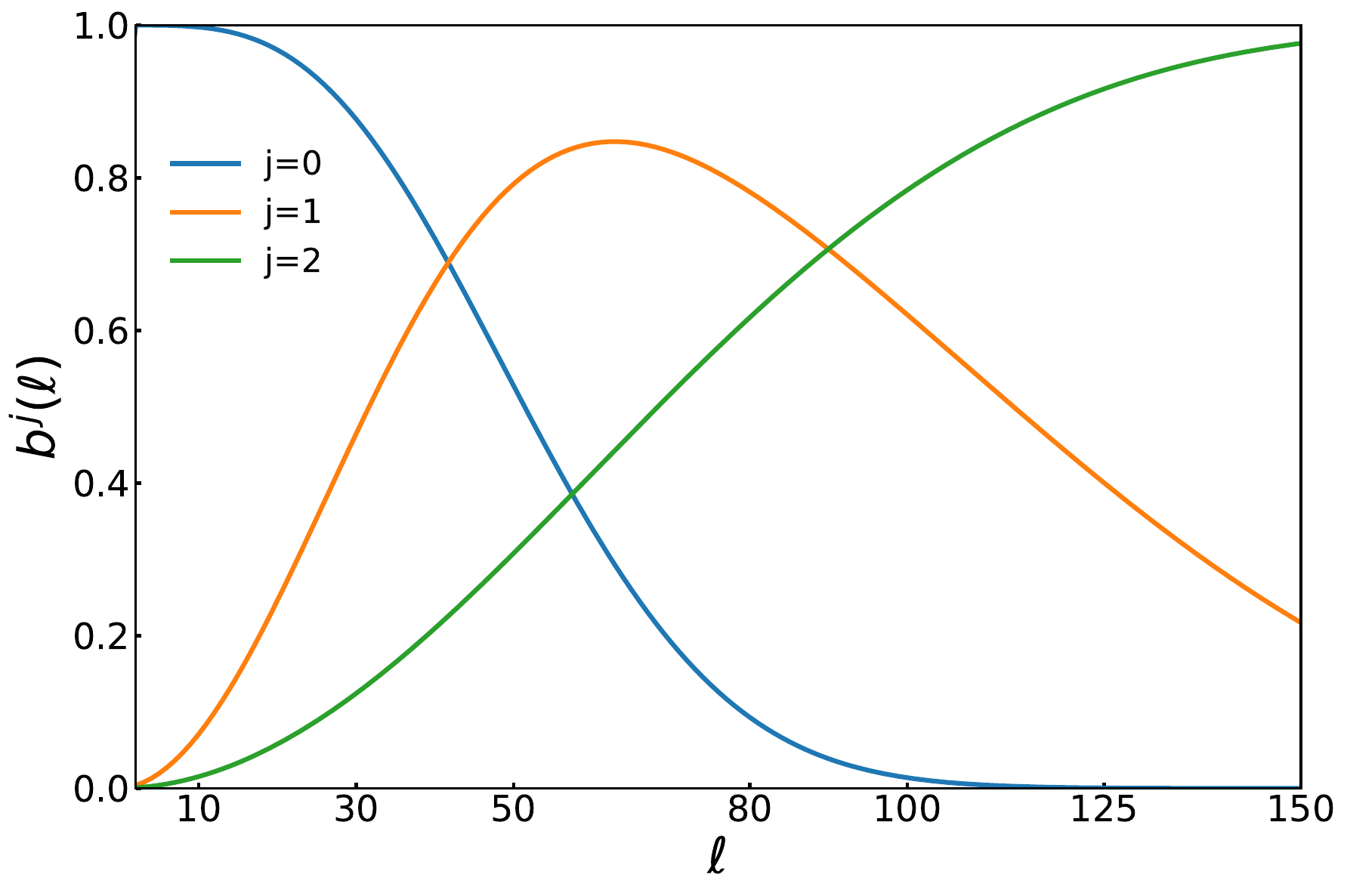}
    \includegraphics[width=0.495\textwidth]{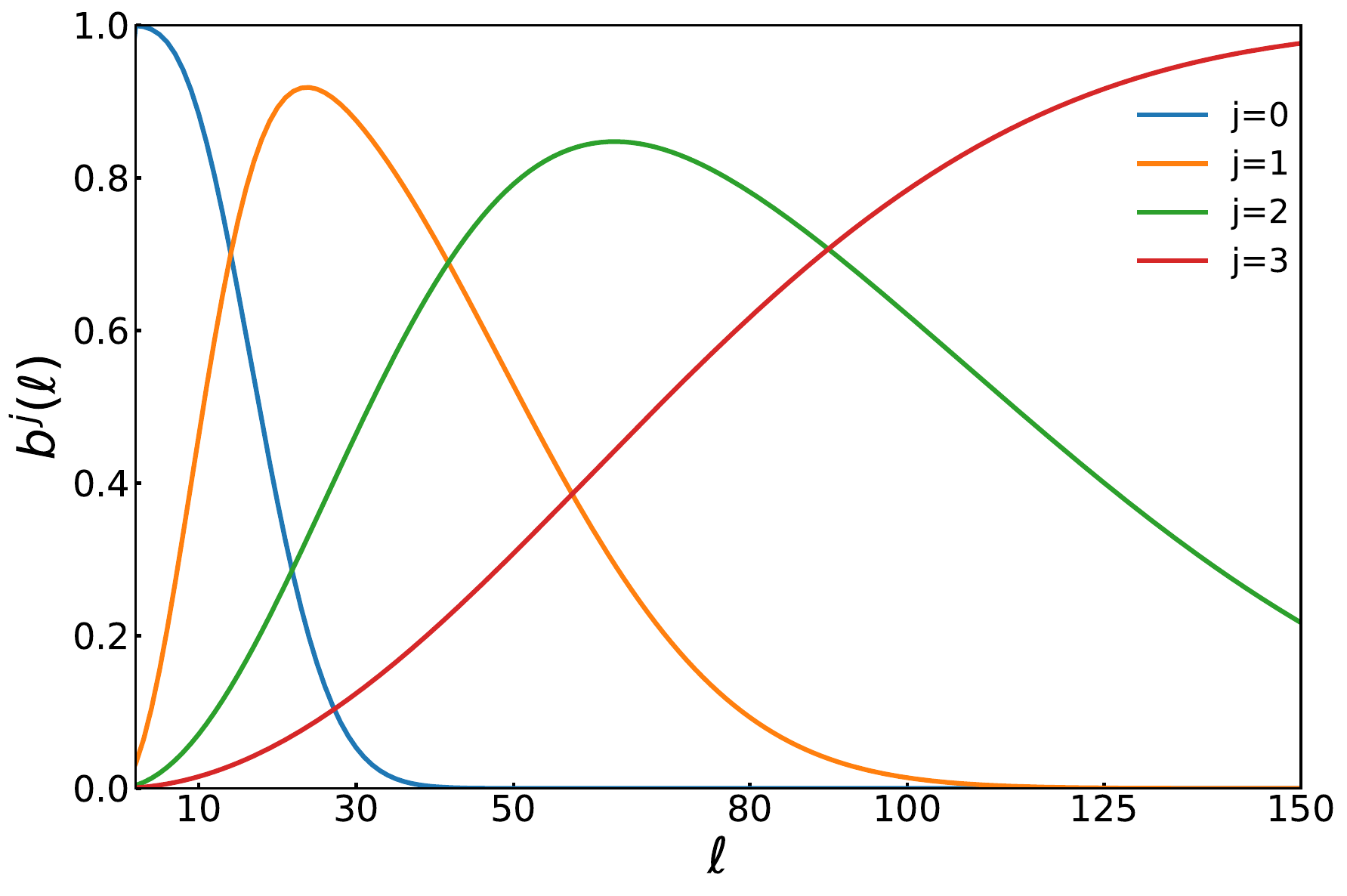}
	\caption{Harmonic needlet filters used in the component separation step (left) and in the GNILC derivation of the multifrequency cleaned foreground maps (right). Both configurations employ mexican-hat needlet bands \cite{mexican_needlets}.}
	\label{fig:needlets}
\end{figure}

To forecast a realistic reconstruction of the CMB $B$-mode polarization pattern from \lb-like simulated maps, we apply a state-of-the-art component separation procedure. Specifically, we implement an advanced ILC technique \cite{ILC}, applied independently at different angular scales taking advantage of needlet filtering of the input data \cite{Marinucci2008}, leading to the so-called Needlet ILC (NILC) \cite{NILC}. The procedure consists of the following steps:

\begin{enumerate}
\item The multifrequency data-set is brought to a common angular resolution, chosen to be the one of the lowest frequency channel in this case: FWHM$\ =70.5\ \text{arcmin}$. $Q$ and $U$ maps are then transformed into $B$-mode maps via a full-sky harmonic decomposition.
\item The $B$-mode maps are convolved with a set of needlet filters $b^{(j)}$ \citep{Marinucci2008}, where the index $j$ labels the different kernels, resulting in multiple multifrequency sets of needlet coefficients $\beta^{(j)}(\hat{n},\nu)$, each sensitive to a specific range of angular scales. The adopted set of needlet bands in harmonic space is shown in the left panel of Figure~\ref{fig:needlets}.
\item For each needlet scale $j$, the multifrequency needlet maps are linearly combined with frequency- and pixel-dependent weights to yield a cleaned estimate of the CMB signal $\tilde{\beta}_{\text{CMB}}^{(j)}(\hat{n})$:
\begin{equation}
\tilde{\beta}_{\text{CMB}}^{(j)}(\hat{n}) = \boldsymbol{w}_{(j)}(\hat{n}) \cdot \beta^{(j)}(\hat{n},\nu)
= \boldsymbol{w}_{(j)}(\hat{n}) \cdot \big(\beta^{(j)}_{\text{CMB}}+\beta^{(j)}_{\text{fgds}}+\beta^{(j)}_{\text{noi}}\big)(\hat{n},\nu),
\label{eq:wilc}
\end{equation}
where the contributions from CMB, foregrounds, and noise are explicitly separated. The weights $\boldsymbol{w}_{(j)}$ are derived so as to minimize the local variance of the solution while preserving the CMB signal, i.e. $\boldsymbol{w}_{(j)}^T(\hat{n}) \cdot A_{\text{CMB}}=1$, with $A_{\text{CMB}}$ denoting the CMB SED in the adopted units \cite{NILC}. The combination coefficients satisfying these conditions are obtained as:
\begin{equation}
    \boldsymbol{w}_{(j)}(\hat{n}) = 
    \left[\mathbf{A}_{\text{CMB}}^{T} \left(C_{(j)}^{-1}(\hat{n})\right) \mathbf{A}_{\text{CMB}}\right]^{-1} 
    \mathbf{A}_{\text{CMB}}^{T} C_{(j)}^{-1}(\hat{n}),
\label{eq:ilc_w}
\end{equation}
where $C_{(j)}$ is the empirical covariance matrix:
\begin{equation}
    (C_{(j)})_{\nu\nu^{\prime}}(\hat{n}) 
    = \left\langle \beta^{(j)}(\hat{n}^{\prime},\nu)\cdot
      \beta^{(j)}(\hat{n}^{\prime},\nu^{\prime}) \right\rangle_{\hat{n}^{\prime}\in \mathcal{D}_{\hat{n}}},
    \label{eq:data_cov}
\end{equation}
estimated as a local average of cross-frequency input needlet maps over a sky domain $\mathcal{D}_{\hat{n}}$. In practice, $\mathcal{D}_{\hat{n}}$ is a circular region centered on the pixel $\hat{n}$, with pixel weights that decrease smoothly and symmetrically toward the borders. The size of each domain depends on the needlet scale $j$: low-$j$ values (corresponding to larger angular scales) require broader domains to include a sufficient number of modes for an accurate covariance estimation.
\item The recovered CMB coefficients from the different needlet scales are finally recombined through an inverse needlet transform \cite{Marinucci2008} to produce a real-space CMB map.
\end{enumerate}

In addition to the baseline NILC implementation, we also consider an alternative methodology. The analysis in \cite{MCNILC} demonstrated that residual foreground and noise contamination in the CMB $B$-mode solution can be further reduced by refining the concept of locality in the variance minimization, moving from a purely spatial basis to one based on foreground properties. Specifically, the sky is partitioned into distinct regions each characterized by similar spectral properties of the $B$-mode foregrounds. The ILC minimization and weight estimation of Equation~\ref{eq:ilc_w} are then implemented independently within each region. This defines the so-called Multi-Clustering NILC (MC-NILC) technique.  

To trace the variability of the foreground spectral properties in a model-independent fashion, an optimal blind tracer has been identified: the ratio of two frequency channels, one dust-dominated (in the considered instrumental configuration, the $337$\,GHz channel) and one with strong constraining power for the CMB reconstruction (here the $119$\,GHz channel) \cite{MCNILC}. This tracer must be reconstructed directly from the observed data. For this purpose, two cleaned templates of the foreground emission—with substantially reduced contributions from CMB and noise—are derived at the relevant frequencies through the application of the GNILC technique (further details of this methodology are provided in Section \ref{ssec:templ}).  

At present, a successful partitioning can only be achieved on large angular scales. Consequently, MC-NILC is applied exclusively to the first needlet band ($j=0$), while standard NILC is used for all other scales. We refer to this hybrid approach as (MC-)NILC.

Both NILC and (MC-)NILC are applied to the input simulations described in Section~\ref{ssec:sims}, making use of dedicated component-separation modules implemented in the \texttt{BROOM} package. The output of these pipelines consists of full-sky $B$-mode maps of the reconstructed CMB signal.

\subsection{Tensor-to-scalar ratio inference}
\label{ssec:spectra}
To infer the posterior distribution of the tensor-to-scalar ratio from the reconstructed CMB signal, we compute the $B$-mode angular power spectrum of the maps, which is then used as input to the cosmological likelihood. Prior to the power spectrum estimation, the reconstructed maps are masked to suppress leakage from the brightest foreground residuals along the Galactic plane. Two masking strategies are considered in this work, corresponding to different retained sky fractions, in order to test the robustness of the proposed "marginalization" procedure. Specifically, we adopt the \texttt{GAL60} and \texttt{GAL40} Galactic masks released by the Planck Collaboration\footnote{\url{https://pla.esac.esa.int/}}, which retain $60\%$ and $40\%$ of the sky, respectively. These masks are constructed from intensity data and may therefore be suboptimal for polarization analyses. However, this actually represents an advantage for our purposes, as it allows us to validate the marginalization approach under extreme conditions, where foreground residuals are expected to be significant and could therefore severely bias the inferred tensor-to-scalar ratio. 

For simplicity, the angular power spectra are computed using the \emph{anafast} routine of the \texttt{healpy} Python package \cite{healpy}. No $E$–$B$ leakage correction is required, since the spectra are computed directly from the masked $B$-mode maps. The loss of power induced by sky masking is corrected by a factor $1/f_{\text{sky}}$, while that due to finite resolution is accounted for by dividing by the corresponding harmonic transfer function. The only effect of masking not explicitly accounted for in this framework is the induced correlation between multipoles. As a consistency check, we repeat the analysis for selected cases using a pseudo-$C_\ell$ correction following the \texttt{MASTER} approach \cite{MASTER, MASTER2}, as implemented in the \texttt{NaMaster} package \cite{pymaster}, and obtain consistent results.

We consider different cases corresponding to all combinations of sky models, input values of $r$, adopted component separation pipelines, and masking strategies. For each case, as described in Section~\ref{ssec:sims}, we have $100$ cleaned CMB maps and their associated angular power spectra ($C_{\ell}^{\text{out}}$). To derive the expected posterior distribution of the tensor-to-scalar ratio, we compute the average spectrum across simulations ($\langle C_{\ell}^{\text{out}}\rangle$) and plug it into the following cosmological inverse-Wishart log-likelihood \citep{Hamimeche2008,gerbino2020}:

\begin{equation}
\log\mathcal{L}(r) = -f_{\textrm{sky}} \frac{2\ell + 1}{2} \left[
\frac{\langle C_{\ell}^{\text{out}} \rangle}{C_{\ell}(r)}
+ \log{C_{\ell}(r)}
- \frac{2\ell - 1}{2\ell + 1} \log{\langle C_{\ell}^{\text{out}} \rangle}
\right],
\label{eq:like}
\end{equation}
where 
\begin{equation}
    C_{\ell}(r) = C_{\ell}^{\text{lens}} + r \cdot C_{\ell}^{\text{tens}}(r=1) + \langle C_{\ell}^{\text{noi}}\rangle
\label{eq:model_1}
\end{equation}
is the theoretical $B$-mode angular power spectrum including lensing, tensor perturbations, and residual instrumental noise. The latter is estimated as the average of the power spectra of noise-only residuals after component separation, obtained by combining noise-only input maps with the component separation weights. The posterior of cosmological parameters (here, only the tensor-to-scalar ratio $r$) is sampled with a Markov Chain Monte Carlo (MCMC) algorithm using the \texttt{emcee} Python package \cite{emcee}, adopting a uniform prior $0 \leq r \leq 0.05$. Although only one parameter is varied, we employ MCMC sampling to ensure consistency with the case where two parameters are sampled (presented in Section \ref{ssec:marginal}). We note that the lensing amplitude is kept fixed, so any residual foreground contamination in the cleaned CMB maps is absorbed as a bias in $r$. 

Strictly speaking, the likelihood in Equation~\ref{eq:like} is exact only for full-sky, signal-dominated power spectra \cite{Hamimeche2008, gerbino2020}. This condition is not met in our case, where sky masking introduces non-trivial correlations among multipoles and residuals can contribute significantly to the total power budget. Nonetheless, we adopt this likelihood for three main reasons: (i) it is straightforward to implement, and the main conclusions are not expected to be sensitive to the exact likelihood formalism; (ii) it allows for direct comparison with similar analyses in the literature \cite{PTEP, MCNILC, puglisi2022}; and (iii) our primary objective is to compare the posteriors of $r$ obtained when foreground residuals are explicitly included in the spectral model with those that neglect them, making the internal consistency of the chosen likelihood the most relevant criterion. Overall, the adoption of this likelihood in this context is expected to yield slightly inaccurate estimates of the error bounds—whose precise determination is not the main objective of this work—and, in some cases, to introduce a bias in the inferred cosmological parameters. However, we anticipate that no statistically significant bias is observed in any of the considered marginalization cases.

\section{Construction of a spectral template of foreground residuals}
\label{sec:marginal}
In this section, we present our novel procedure for deriving a cleaned template of the foreground residuals (Section~\ref{ssec:templ}) and describe how its angular power spectrum is incorporated into the cosmological likelihood to debias the tensor-to-scalar ratio (Section~\ref{ssec:marginal}).

\subsection{Derivation of the template of the foreground residuals}
\label{ssec:templ}
The key step in deriving a $B$-mode template of foreground residuals is to obtain cleaned estimates of the Galactic foreground emission across the observed frequency range. These foreground $B$-mode maps, with reduced contamination from CMB and instrumental noise, are then combined using the same component separation weights employed for CMB recovery in Equation~\ref{eq:wilc}. This procedure naturally produces a cleaned estimate of the distribution of foreground residuals that remain in the reconstructed CMB solution after component separation. In the proposed approach, the multifrequency Galactic emission data set is obtained by applying the GNILC technique \cite{GNILC, GNILC_intro} to the simulated microwave maps. Although GNILC has been extensively described in the literature, we briefly summarize here the key steps required to successfully implement the pipeline, given that it constitutes the core methodology of our procedure.

% The derivation of a cleaned template of the foreground residuals is detailed below.

\texttt{Step 1} and \texttt{Step 2} of the component separation procedure (see Section~\ref{ssec:compsep}) are first applied to the input $Q/U$ maps. The adopted set of needlet bands in this case is shown in the right panel of Fig.~\ref{fig:needlets}.

The GNILC technique is then applied to the pre-processed $B$-mode needlet maps. At each needlet scale, the following steps are carried out:

\begin{enumerate}
\item Two covariance matrices are computed: one from the data as in Equation \ref{eq:data_cov},
\begin{equation}
(C_{(j)})_{\nu\nu'}(\hat{n})=\langle \beta^{(j)}(\hat{n}^{\prime},\nu) \cdot \beta^{(j)}(\hat{n}^{\prime},\nu') \rangle_{\hat{n}^{\prime}\in \mathcal{D}_{\hat{n}}}
= (C^{\text{CMB}}_{(j)})_{\nu\nu'}(\hat{n}) + (F_{(j)})_{\nu\nu'}(\hat{n}) + (N_{(j)})_{\nu\nu'}(\hat{n}),
\label{eq:covariances}
\end{equation}
and one from realistic simulations of the nuisance components to be deprojected (CMB and noise in this case):
\begin{equation}
(\tilde{N}_{(j)})_{\nu\nu'}(\hat{n})=(C^{\text{CMB}}_{(j)})_{\nu\nu'}(\hat{n}) + (N_{(j)})_{\nu\nu'}(\hat{n}),
\label{eq:covariances_1}
\end{equation}
where $C^{\text{CMB}}_{(j)}$, $F_{(j)}$, and $N_{(j)}$ denote the CMB, foreground, and noise covariance contributions, respectively. For simplicity, in the following we omit the $(j)$ and $\nu\nu'$ subscripts.
\item The \emph{whitened} data covariance is then constructed:  
\begin{equation}
    \tilde{C}(\hat{n}) = \big( \tilde{N}^{-1/2} C \tilde{N}^{-1/2} \big)(\hat{n}) 
    \simeq \big(\tilde{N}^{-1/2} F \tilde{N}^{-1/2}\big)(\hat{n}) + \mathds{1}\,,
    \label{eq:whitened_covar}
\end{equation}
whose diagonalization yields  
\begin{equation}
    \tilde{C}(\hat{n}) \simeq 
    \begin{pmatrix} U_{F}(\hat{n})\ |\ U_{N}(\hat{n}) \end{pmatrix}
    \begin{pmatrix}
        \lambda_{1}(\hat{n}) & & & & & \\
        & \ddots & & & & \\
        & & \lambda_{m(\hat{n})}(\hat{n}) & & & \\
        \hline
        & & & 1 & & \\
        & & & & \ddots & \\
        & & & & & 1
    \end{pmatrix}
    \begin{pmatrix}
        U_{F}(\hat{n})^T \\
        \hline
        U_{N}(\hat{n})^T
    \end{pmatrix}\,,
    \label{eq:C_tilde}
\end{equation}
where eigenvalues significantly larger than unity ($\lambda_{1},\dots,\lambda_{m}$) correspond to statistically independent Galactic emission modes that dominate over the CMB and instrumental noise variance at a given sky position $\hat{n}$. These foreground modes are represented by the set of eigenvectors $U_{F}(\hat{n})$. The rank of the foreground sub-space $m(\hat{n})$, i.e. the number of eigenvalues significantly departing from $1$, is derived by minimizing the Akaike Information Criterion (AIC) \cite{GNILC}:
\begin{equation}
    \mathrm{AIC}(m,\hat{n}) = 2m +  \sum_{k=m+1}^{N_{\text{freq}}} \left[ \lambda_k(\hat{n}) - \log \lambda_k(\hat{n}) - 1 \right],
\label{eq:AIC}
\end{equation}
with $N_{\text{freq}}$ the number of observed frequency channels.
\item Finally, the weights that reconstruct the Galactic emission across all frequency channels are obtained by deprojecting eigenmodes with eigenvalues close to unity in Equation~\ref{eq:C_tilde} (i.e., dominated by CMB and noise) \cite{GNILC, GNILC_intro}:  
\begin{equation}
    W(\hat{n}) = \big[\tilde{F}\,\big(\tilde{F}^{T}C^{-1}\tilde{F}\big)^{-1}\tilde{F}^{T}C^{-1}\big](\hat{n}),
    \label{eq:gnilc_w}
\end{equation}
with $\tilde{F} = \tilde{N}^{1/2}U_{F}$. The weight matrix $W$ has dimension $(N_{\text{freq}} \times N_{\text{freq}} \times N_{\text{pix}})$, where $N_{\text{pix}}$ denotes the number of pixels. Applying $W$ to the input needlet maps with dimension $(N_{\text{freq}} \times N_{\text{pix}})$ yields a multifrequency set of cleaned foreground $B$-mode maps. 

\end{enumerate}

For each frequency channel, the cleaned foreground needlet maps at the different scales are then recombined through an inverse needlet transformation \cite{Marinucci2008}, following the same procedure adopted in \texttt{Step 4} of the component-separation pipeline (see Section~\ref{ssec:compsep}). This final step enables the full-scale reconstruction of the Galactic emission.

The GNILC procedure is robust against foreground uncertainties, as no explicit modeling of Galactic emission is implemented or required. On the other hand, the GNILC technique implicitly assumes a perfect characterization of the two-point statistics of the contaminants to be deprojected, since the nuisance covariance $\tilde{N}$ is a required input. In practice, this assumption is quite robust: the expected CMB power spectrum is very well known and the instrumental noise properties can be characterized with high accuracy. The only subtle uncertainty in the $B$-mode implementation arises from the contribution of primordial tensor modes. In this analysis, we do not explore the impact of potential mismatches between the true tensor contribution and the one assumed in the nuisance covariance. However, we do not expect this to have any significant impact because: (i) tensor modes, given the current upper limits on their amplitude \citep{tristram2022, Galloni2023}, contribute appreciably to the CMB $B$-mode covariance only on the largest angular scales, which are never considered in isolation within the first needlet scale; and (ii) even if tensor modes are not explicitly included, they are effectively deprojected together with the lensing $B$-modes, as both share the same frequency scaling.

The cleaned maps of $B$-mode foreground emission, already homogenized to a common angular resolution of FWHM$=70.5\ \text{arcmin}$, are processed in exactly the same way as the input total maps within the CMB reconstruction pipeline. Specifically, they are convolved with the set of needlet filters used in \texttt{Step 2} of the component separation procedure (shown in the left panel of Fig.~\ref{fig:needlets}). The resulting needlet maps represent our best estimate of the foreground contribution to the input needlet maps used in component separation. At each needlet scale, these maps are combined with the corresponding component separation weights $\boldsymbol{w}_{(j)}(\hat{n})$ as done for input data in Equation \ref{eq:wilc}, yielding a scale-specific estimate of the residual foreground contamination in the reconstructed CMB solution. Finally, the residual foreground template in pixel space, $\tilde{f}_{\text{res}}(\hat{n})$, is obtained through an inverse needlet transformation. This procedure, like the component-separation routines, is included in the \texttt{BROOM} package.

For each simulation, an independent set of $100$ multifrequency realizations containing only instrumental noise is first propagated through the GNILC pipeline and then combined with the component-separation weights, following the same procedure used for the GNILC templates. The power spectrum of the resulting \emph{noise residuals}, $C_{\ell}^{\tilde{n}}$, provides an estimate of the noise bias in the angular power spectrum of the residuals template.

To decouple the outcomes of the needlet component separation used for CMB reconstruction (NILC) and for foreground estimation (GNILC), we adopt different hyperparameters for the two pipelines. Specifically, we use distinct sets of needlet bands (as shown in the two panels of Fig.~\ref{fig:needlets}) and different values of the residual ILC bias (see Appendix~A of \cite{NILC} for further details), which effectively control the size of the domains $\mathcal{D}_{\hat{n}}$ over which covariances are independently computed (see Equations~\ref{eq:data_cov} and \ref{eq:covariances}). In practice, lower values of the ILC bias correspond to larger circular domains.

Although not fully explored in this work, the GNILC implementation can be further modified to improve the reconstruction of the foreground-residual template. One relevant hyperparameter is the rank of the Galactic subspace, $m(\hat{n})$ (see Equation~\ref{eq:C_tilde}), which determines the number of modes retained in the signal subspace. In practice, one can choose to preserve more (or fewer) modes than those indicated by the AIC criterion in Equation~\ref{eq:AIC}, thereby achieving a better (or poorer) reconstruction of the foreground emission at the cost of enhanced (or reduced) contamination from noise and CMB. A second hyperparameter available to the user is the possibility to enforce a specific level of CMB deprojection. This corresponds to constraining the GNILC weights in Equation~\ref{eq:gnilc_w} to remove a fixed fraction of the CMB signal at all frequency channels. However, imposing a stronger deprojection of the CMB signal can, in turn, increase the reconstruction noise due to the additional constraint applied in the weight estimation. Both the rank of the foreground subspace and the level of CMB deprojection can be manually adjusted when applying this methodology within \texttt{BROOM} and may assume different values for distinct needlet bands. In this work, we configure GNILC to preserve $m(\hat{n}) + 1$ independent modes in its reconstruction and to fully deproject the CMB signal at all needlet scales except for $j = 0$ (corresponding to the largest angular scales). This configuration was found to be robust across all considered scenarios.

Finally, we note that, although this work focuses on propagation through the NILC and MC-NILC pipelines, the multifrequency set of foreground templates obtained from the GNILC pipeline can be combined with the weights from any map-based component-separation method—either parametric or blind—implemented in any domain, including pixel, harmonic, or needlet space.

% If NILC and GNILC were implemented with identical hyperparameters, the GNILC-reconstructed signal would coincide with the component deprojected by NILC, and combining GNILC templates with NILC weights would therefore yield a null template.

\subsection{Spectral model}
\label{ssec:marginal}
The spectral template of the foreground residuals, $C_{\ell}^{f_{res}}$, contaminating $C_{\ell}^{\text{out}}$, is derived for each simulation as
\begin{equation}
C_{\ell}^{f_{res}} = C_{\ell}^{\tilde{f}_{res}} - \langle C_{\ell}^{\tilde{n}} \rangle,
\label{eq:cl_templ}
\end{equation}
where $C_{\ell}^{\tilde{f}_{res}}$ is the angular power spectrum of the foreground residuals template obtained with the procedure described in Section \ref{ssec:templ}, and $\langle C_{\ell}^{\tilde{n}} \rangle$ is the average over 100 simulations of the corresponding noise contribution. This second term is required to debias the spectral template from reconstruction noise. This correction is robust, as the debiasing can be performed either by estimating the noise bias from realistic instrumental simulations (as done in this work), or by computing cross-spectra of the map-based templates obtained from different data splits with uncorrelated noise. All spectra are computed using the same routine and masking strategies as described in Section~\ref{ssec:spectra}. The derivation of the tensor-to-scalar ratio posterior is then modified. While the likelihood function and sampling procedure remain unchanged, the power spectrum model of Equation~\ref{eq:like} is replaced by:

\begin{equation}
C_{\ell}(r, A_{f}) = C_{\ell}^{\text{lens}} + r \cdot C_{\ell}^{\text{tens}}(r=1) + \langle C_{\ell}^{\text{noi}}\rangle + A_{f}\cdot \langle C_{\ell}^{f_{res}} \rangle,
\label{eq:model_2}
\end{equation}

so that the MCMC now samples the two-dimensional posterior $P(r, A_{f})$. The additional parameter $A_{f}$ quantifies any mismatch between the amplitude of the actual foreground residuals and that of the derived template in Equation~\ref{eq:cl_templ}. By construction, if the spectral shape of $C_{\ell}^{f_{\text{res}}}$ matches that of the actual foreground residuals, $A_{f}$ absorbs their contribution to $C_{\ell}^{\text{out}}$, thereby mitigating the bias that would otherwise affect the inferred tensor-to-scalar ratio. In this case, we impose the prior $0 \leq A_{f} < \infty$.

\begin{figure}
	\centering
	\includegraphics[width=1.\textwidth]{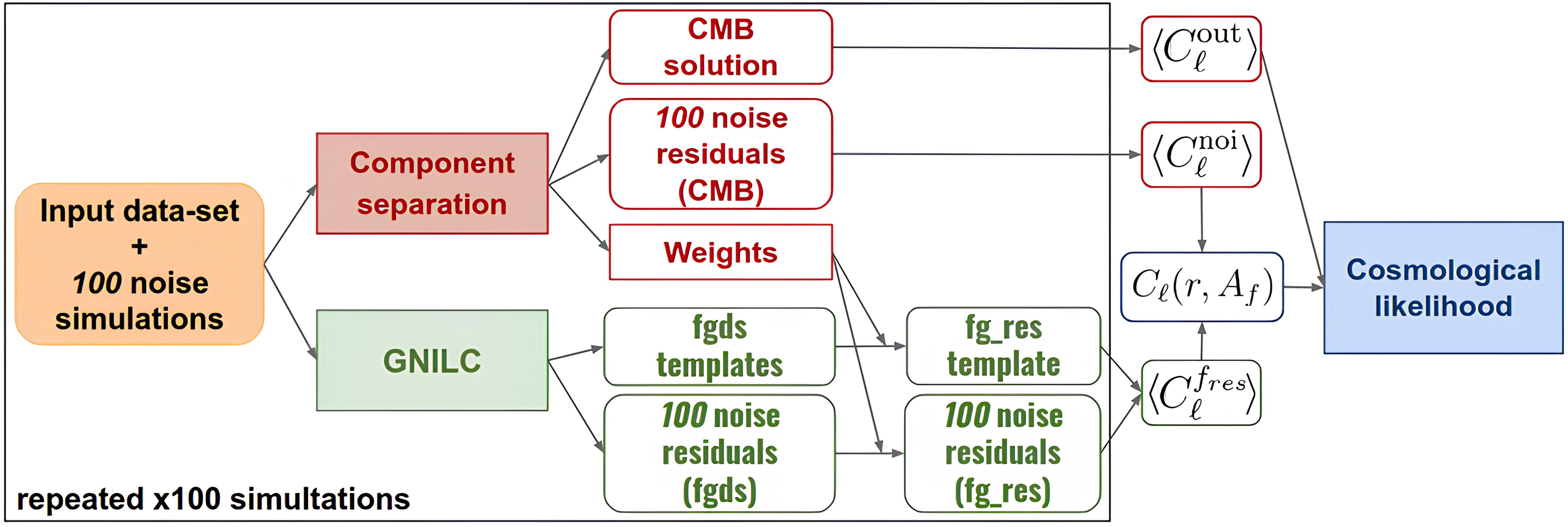}
	\caption{Flowchart of the pipeline used to process each simulated data set, leading to the derived average observed and model angular power spectra, which are then used as inputs to the cosmological likelihood in Equation~\ref{eq:like}.}
	\label{fig:chart}
\end{figure}

A schematic flowchart summarizing the full procedure proposed and implemented in this work, as applied to each simulation, is shown in Figure~\ref{fig:chart}.

% The final posterior of the tensor-to-scalar ratio is then obtained by a naive marginalization of the $2D$ posterior:
% \begin{equation}
%     \tilde{P}(r) = \frac{\int P(r, A_{f}) dA_{f}}{\int P(r, A_{f}) drdA_{f}}
% \end{equation}

\section{Results}
\label{sec:results}
As outlined in Section~\ref{ssec:sims}, we consider six sets of input microwave simulations, corresponding to three values of the tensor-to-scalar ratio ($r = 0,\ 0.004,\ 0.01$) combined with two foreground sky models (\texttt{d1s1} and \texttt{d10s5}). For each set, 100 independent realizations are generated assuming a \lb-like instrumental configuration. These simulations are then processed with two component separation pipelines, NILC and (MC-)NILC (see Section~\ref{ssec:compsep}), and the resulting power spectra are computed using two masking strategies (\texttt{GAL60} and \texttt{GAL40}). Altogether, this amounts to 24 analysis cases. For each case, we report constraints on the tensor-to-scalar ratio, both with and without inclusion of the spectral template of foreground residuals in the model of Equation~\ref{eq:like}.

We begin by comparing the power spectrum of the actual foreground residuals with that of the corresponding template, derived according to Equation~\ref{eq:cl_templ} using both NILC and (MC-)NILC component separation routines. This comparison is, of course, only possible when dealing with simulations rather than real data, but it provides a robust validation test for the proposed methodology. For each simulation and case, a map of the foreground residuals is obtained by processing the input foreground-only maps in the same way as the total data (where all components have been co-added) and by combining them with the component-separation weights of Equation~\ref{eq:ilc_w}.  
\begin{figure}
	\centering
	\includegraphics[width=0.495\textwidth]{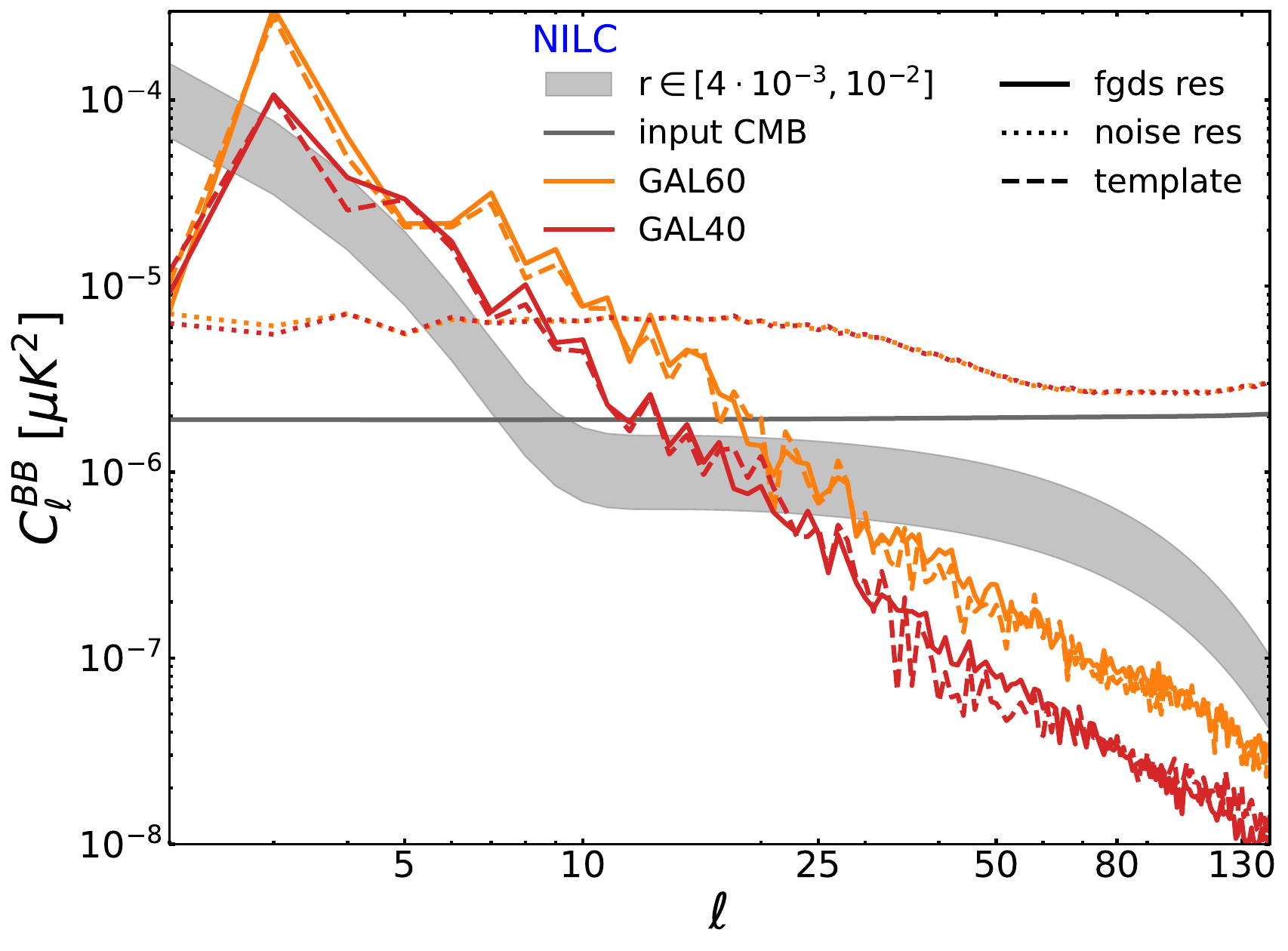}
    \includegraphics[width=0.495\textwidth]{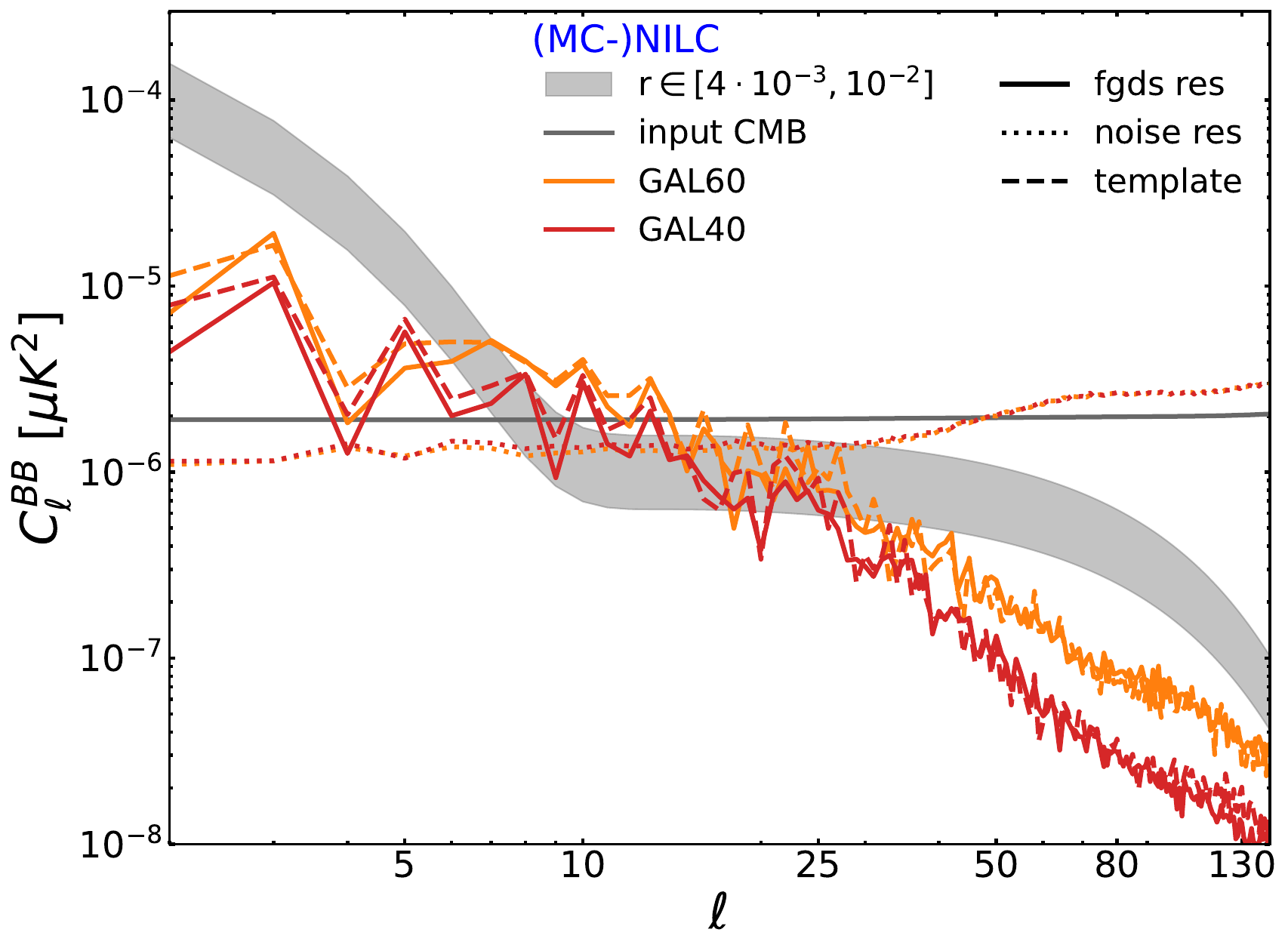}
	\caption{Angular power spectra, averaged over $100$ realizations, of the noise residuals (dotted lines), foreground residuals (solid lines) and denoised template of foreground residuals (dashed lines, obtained with Equation \ref{eq:cl_templ}). Left and right panels report results for NILC and (MC-)NILC, respectively. Orange and red lines correspond to results obtained when \texttt{GAL60} and \texttt{GAL40} masks are used to compute angular power spectra. The reported results refer to the case where input $r=0$. The grey shaded area denotes the range of primordial tensor CMB $B$-mode power spectra corresponding to tensor-to-scalar ratio values $r\in [0.004,0.01]$.}
	\label{fig:spectra_comp}
\end{figure}
\begin{figure}
	\centering
	\includegraphics[width=0.325\textwidth]{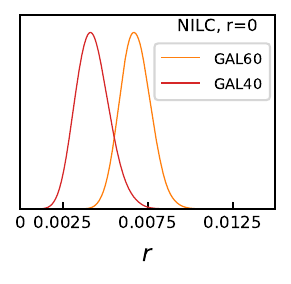}
    \includegraphics[width=0.325\textwidth]{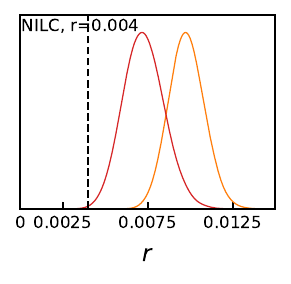}
    \includegraphics[width=0.325\textwidth]{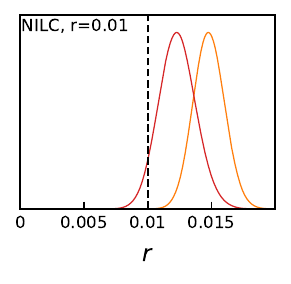}

    \includegraphics[width=0.325\textwidth]{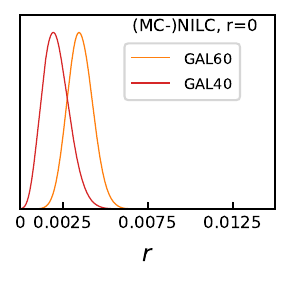}
	\includegraphics[width=0.325\textwidth]{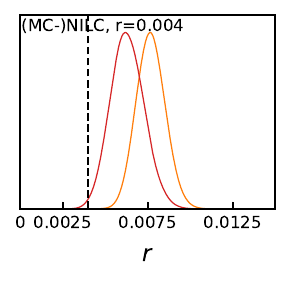}
    \includegraphics[width=0.325\textwidth]{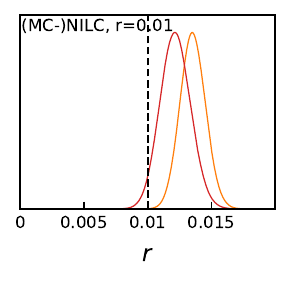}
    \caption{Posterior distributions of the tensor-to-scalar ratio derived from the average observed angular power spectrum after applying NILC (top) and (MC-)NILC (bottom) to simulated data sets with the \texttt{d1s1} foreground model and input CMB signals with $r=0$ (left), $r=0.004$ (center), and $r=0.01$ (right). In these cases, no contribution from foreground residuals is included in the spectral model, as defined in Equation~\ref{eq:model_1}. The vertical black dashed lines indicate the input nonzero values of the tensor-to-scalar ratio.}
	\label{fig:posteriors_nomarg}
\end{figure}

Figure~\ref{fig:spectra_comp} presents the average spectra over all simulations, computed with the \texttt{GAL60} and \texttt{GAL40} masks. As an illustrative example, we focus here on the case based on the \texttt{d1s1} model with $r=0$ as input. A very good agreement is found between the spectral shape of the power spectrum of the actual foreground residuals and that of the template across the full multipole range, for both component-separation methods and both masking strategies. The same conclusions hold for the additional cases considered but not shown here. As discussed in Section~\ref{ssec:marginal}, the agreement in spectral shape is the most relevant aspect, since any mismatch in amplitude can be accounted for by the additional parameter $A_{f}$ in Equation~\ref{eq:model_2}.
\begin{figure}
	\centering
	\includegraphics[width=0.325\textwidth]{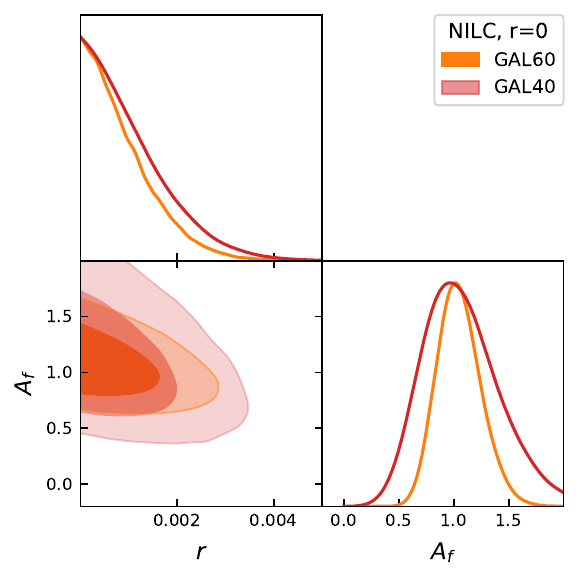}
    \includegraphics[width=0.325\textwidth]{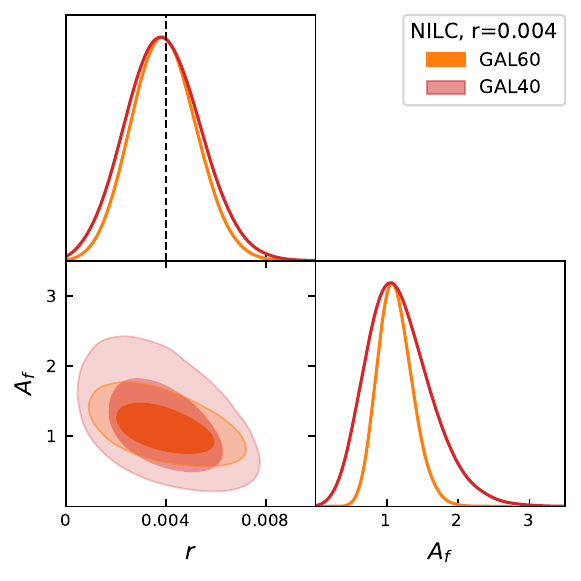}
    \includegraphics[width=0.325\textwidth]{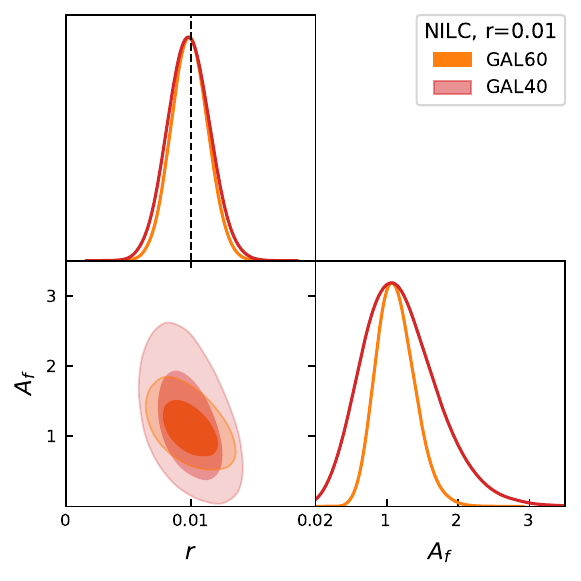}
    \includegraphics[width=0.325\textwidth]{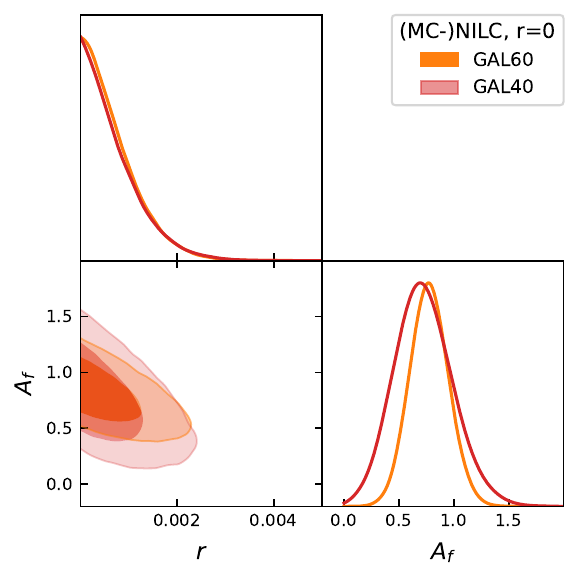}
    \includegraphics[width=0.325\textwidth]{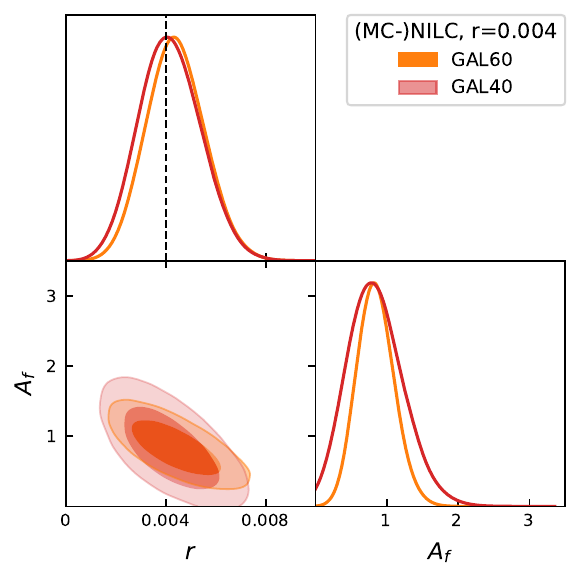}
    \includegraphics[width=0.325\textwidth]{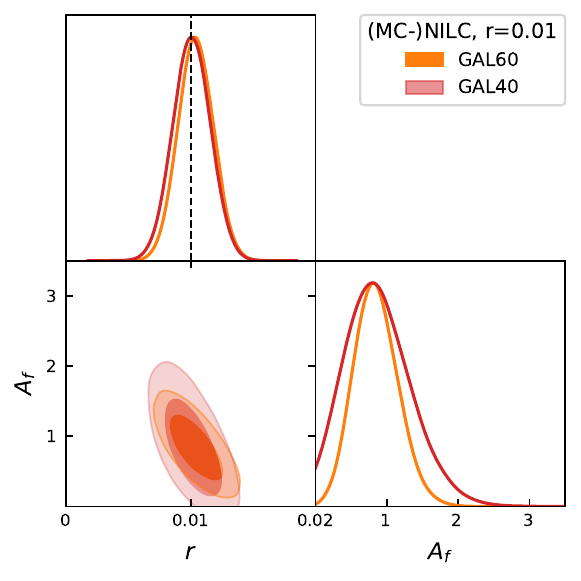}
    \caption{Two-dimensional and one-dimensional posteriors ($r$ and $A_{f}$, shown in the top and right sub-panels of each panel, respectively) obtained from sampling the likelihood of Equation~\ref{eq:like} when the spectral template of foreground residuals is included in the model of Equation~\ref{eq:model_2}. The top and bottom rows correspond to the application of NILC and (MC-)NILC, respectively, to the simulated \texttt{d1s1} data set. Different input values of the tensor-to-scalar ratio are shown: $r=0$ (left), $r=0.004$ (center), and $r=0.01$ (right). The vertical black dashed lines indicate the input nonzero values of the tensor-to-scalar ratio.}
	\label{fig:posteriors_marg}
\end{figure}

We first propagate the observed power spectrum, obtained after component separation, into the cosmological likelihood without including any foreground contribution in the spectral model as in Equation~\ref{eq:model_1}. The resulting posteriors of $r$ for NILC and (MC-)NILC, obtained for both masking strategies and all considered input values of $r$, are shown in Figure~\ref{fig:posteriors_nomarg}. As in Figure~\ref{fig:spectra_comp}, we report results for the \texttt{d1s1} scenario as a representative example.
In all cases, we find that foreground residuals significantly bias the best-fit value of $r$ relative to the input value in the simulations. As expected, the use of the more aggressive \texttt{GAL40} mask and/or of the (MC-)NILC methodology partially mitigates this bias. This improvement arises, respectively, from the exclusion of broader regions close to the Galactic plane and from the more effective subtraction of foreground contaminants on large angular scales achieved by the MC-NILC framework (see Section~\ref{ssec:compsep}).
In all cases shown in Figure~\ref{fig:posteriors_nomarg}, the input fiducial value $r_{\text{in}}$ lies outside the $95\%$ confidence region, indicating a statistically significant level of contamination.

We then re-sample the cosmological likelihood of Equation~\ref{eq:like}, this time including the spectral template derived in Equation~\ref{eq:cl_templ} within the model of Equation~\ref{eq:model_2}. The resulting posterior is two-dimensional, with parameters $r$ and $A_{f}$, where $A_{f}$ parametrizes the relative amplitude of the template with respect to the actual foreground residuals in the observed power spectrum. The choice of introducing a simple amplitude rescaling factor $A_{f}$ in Equation~\ref{eq:model_2} to account for potential mismatches between the actual foreground residual contamination and the derived template represents the most agnostic and minimal assumption in the spectral model. This choice is strongly supported by the excellent agreement in the spectral shapes of the two angular power spectra shown in Figure~\ref{fig:spectra_comp}. 

For the same illustrative cases as those shown without marginalization in Figure~\ref{fig:posteriors_nomarg}, the 2D and 1D posterior distributions, obtained  when the spectral model of the foreground residuals is included, are presented in Figure~\ref{fig:posteriors_marg}.
As expected, we find a clear detection of foreground residuals, since $A_{f}$ is not statistically compatible with zero in any case. This, in turn, yields an unbiased estimate of the tensor-to-scalar ratio, thus effectively fully mitigating the impact of foreground residuals into cosmological inference of the amplitude of primordial $B$ modes. The result proves robust, holding across different masking strategies, component separation approaches, and input tensor amplitudes. The best-fit value of $A_{f}$ remains stable across the two masking strategies for all the cases considered, demonstrating the correct rescaling of the templates with the adopted mask, in agreement with that observed in the actual foreground residuals. At the same time, the marginalization procedure retains its effectiveness in fully mitigating the foreground-induced bias across all cases. Interestingly, we note that the correlation pattern between the two fitted parameters varies with the adopted amplitude of the input tensor perturbations ($r_{\text{in}}$), and, to a lesser extent, with the employed masking strategy.
\begin{table}[t]
\centering
\makebox[\textwidth][c]{\begin{tabular}{lccc}
\hline
 & $r_{\text{in}}=0$ & $r_{\text{in}}=0.004$ & $r_{\text{in}}=0.01$ \\
\hline
NILC - \texttt{d1s1} - \texttt{GAL60} & $0.0067\pm0.0009$ & $0.01^{+0.0011}_{-0.001}$ & $0.015\pm0.0012$ \\
NILC - \texttt{d1s1} - \texttt{GAL60} (marg.) & $<0.00106$ & $0.0039\pm0.0013$ & $0.01^{+0.0015}_{-0.0014}$ \\
NILC - \texttt{d1s1} - \texttt{GAL40} & $0.0042^{+0.0010}_{-0.0009}$ & $0.0072^{+0.0013}_{-0.0012}$ & $0.0125^{+0.0015}_{-0.0014}$ \\
NILC - \texttt{d1s1} - \texttt{GAL40} (marg.) & $<0.00128$ & $0.0039\pm0.0015$ & $0.01\pm0.0017$ \\
(MC-)NILC - \texttt{d1s1} - \texttt{GAL60} & $0.0035^{+0.0008}_{-0.0007}$ & $0.0077^{+0.0009}_{-0.0008}$ & $0.0135\pm0.001$ \\
(MC-)NILC - \texttt{d1s1} - \texttt{GAL60} (marg.) & $<0.00083$ & $0.0044^{+0.0012}_{-0.0011}$ & $0.0104\pm0.0014$ \\
(MC-)NILC - \texttt{d1s1} - \texttt{GAL40} & $0.0021^{+0.0008}_{-0.0007}$ & $0.0063\pm0.001$ & $0.012\pm0.0012$ \\
(MC-)NILC - \texttt{d1s1} - \texttt{GAL40} (marg.) & $<0.00084$ & $0.0041^{+0.0013}_{-0.0012}$ & $0.0101\pm0.00015$\\
NILC - \texttt{d10s5} - \texttt{GAL60} & $0.0084^{+0.0013}_{-0.0012}$ & $0.012\pm0.0014$ & $0.0173\pm0.0016$ \\
NILC - \texttt{d10s5} - \texttt{GAL60} (marg.) & $<0.002$ & $0.004^{+0.0021}_{-0.002}$ & $0.01\pm0.0023$ \\
NILC - \texttt{d10s5} - \texttt{GAL40} & $0.0065^{+0.0015}_{-0.0014}$ & $0.01^{+0.0017}_{-0.0016}$ & $0.0153\pm0.0019$ \\
NILC - \texttt{d10s5} - \texttt{GAL40} (marg) & $<0.0023$ & $0.004^{+0.0023}_{-0.0022}$ & $0.0098\pm0.0028$ \\
(MC-)NILC - \texttt{d10s5} - \texttt{GAL60} & $0.0056^{+0.001}_{-0.0009}$ & $0.01\pm0.001$ & $0.016^{+0.0012}_{-0.0011}$ \\
(MC-)NILC - \texttt{d10s5} - \texttt{GAL60} (marg) & $<0.0011$ & $0.0044^{+0.0017}_{-0.0016}$ & $0.01^{+0.0025}_{-0.0024}$ \\
(MC-)NILC - \texttt{d10s5} - \texttt{GAL40} & $0.0045\pm0.0011$ & $0.0088\pm0.0012$ & $0.015\pm0.0014$ \\
(MC-)NILC - \texttt{d10s5} - \texttt{GAL40} (marg.) & $<0.0012$ & $0.0043^{+0.0019}_{-0.0018}$ & $0.01 \pm 0.0026$ \\
\hline
\end{tabular}}
\caption{Constraints on the tensor-to-scalar ratio derived from the posteriors obtained by sampling the likelihood in Equation~\ref{eq:like}. Cases where a spectral template of foreground residuals is included in the model (Equation~\ref{eq:model_2}) are marked with “(marg.)”. When the null value does not lie within the $68\%$ confidence interval, a detection is reported as the median value with its corresponding $68\%$ confidence interval; otherwise, an upper bound at $68\%$ confidence level is provided.}
\label{tab:r_posts}
\end{table}

The summary statistics of the $r$ posteriors, obtained with and without including the template of foreground residuals in the spectral model, are reported in Table~\ref{tab:r_posts} for all the analysed 24 cases, including those based on the \texttt{d10s5} scenario. We find that in every case foreground residuals affect the $r$ estimation, leading to significant biases in the tensor-to-scalar ratio. Once these residuals are properly accounted for, we recover an unbiased upper bound for $r=0$ or a consistent detection when the input $r$ is nonzero. We note that the debiasing of the tensor-to-scalar ratio is accompanied, as expected, by an increase in uncertainty, stemming from the additional parameter introduced in the model and sampled in the MCMC analysis. From this perspective, achieving a good match between the spectral features of the actual foreground residuals and those of the template is crucial, as it minimizes the number of nuisance parameters that must be included in the spectral model of Equation~\ref{eq:model_2}, thereby reducing the impact on the uncertainty of the inferred cosmological parameters.

An overall comparison between the $r$ constraints obtained for the \texttt{GAL60} and \texttt{GAL40} cases, when the foreground spectral model is included, suggests that in terms of overall sensitivity, it is preferable to retain a larger sky fraction and thus marginalize over a correspondingly larger level of contamination rather than exclude it. This conclusion, however, is expected to hold as long as there is a good agreement between the template and the actual foreground residuals. The gain in sensitivity achieved by retaining a larger sky fraction—at the cost of a higher residual amplitude to be marginalized over—is actually underestimated by the likelihood framework adopted in this work. Indeed, every term included in the model of the inverse-Wishart likelihood in Equation~\ref{eq:like} contributes as additional variance in the parameter inference. This applies, for instance, to the enhanced power from the foreground residuals absorbed by the term $A_{f}\cdot\langle C_{\ell}^{f_{\mathrm{res}}}\rangle$ in Equation~\ref{eq:model_2}. Yet this effect arises from the use of an idealized likelihood: in practice, such a contribution is not statistical but almost entirely systematic, and thus should not translate into a corresponding increase in the parameter uncertainties. Consequently, in a more realistic likelihood framework, the impact of the larger foreground residual contribution on the spectrum variance would be smaller, further enhancing the difference between optimal marginalization with a larger retained sky fraction and the opposite case.

The size of the bias on the cosmological parameter naturally depends on the chosen component separation method and masking strategy. However, since the procedure is fully model-independent, it provides a general and robust basis for treating potential foreground residual contamination in future reconstructions of the CMB polarization field.

\begin{figure}
	\centering
	\includegraphics[width=0.495\textwidth]{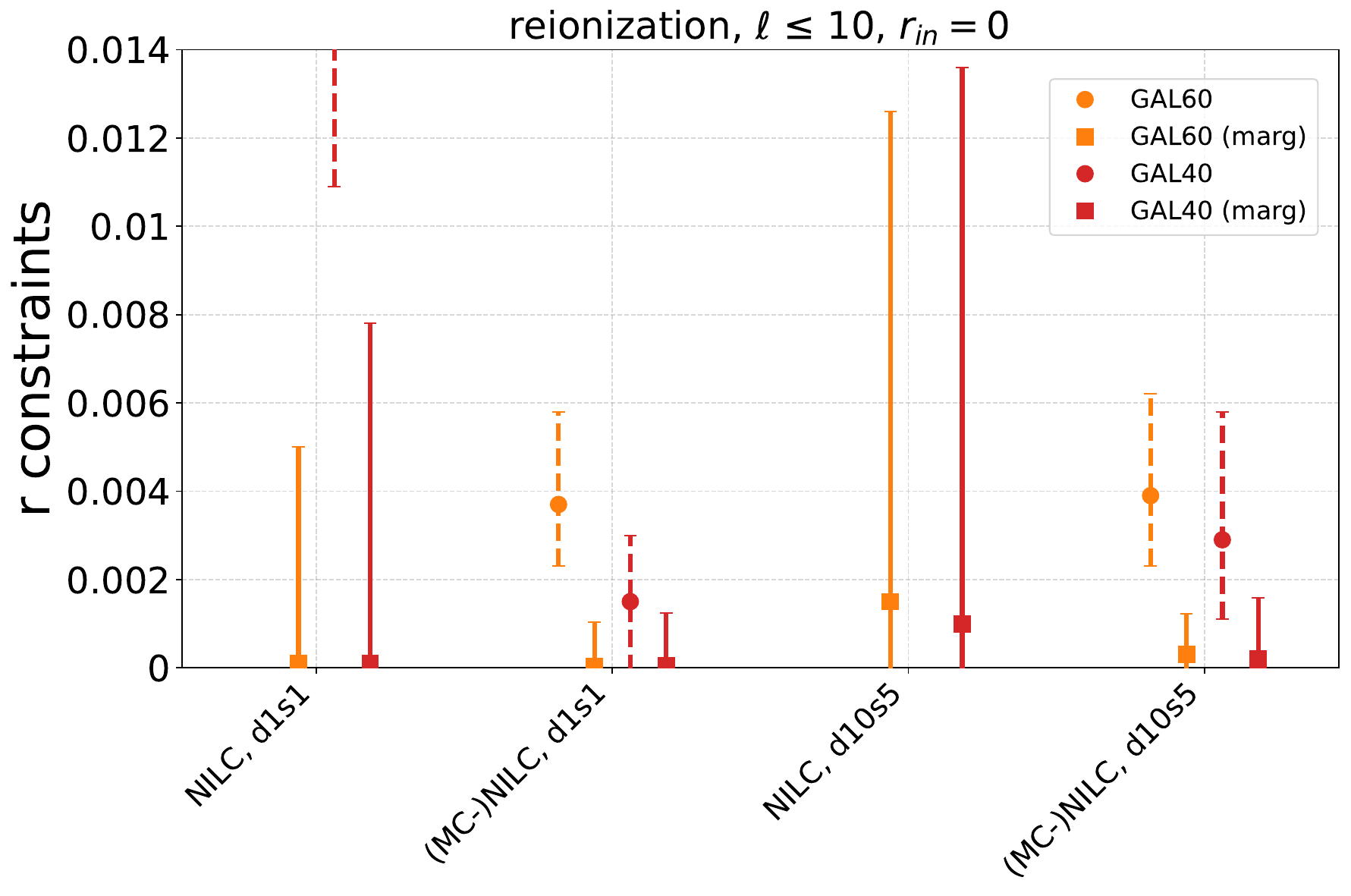}
    \includegraphics[width=0.495\textwidth]{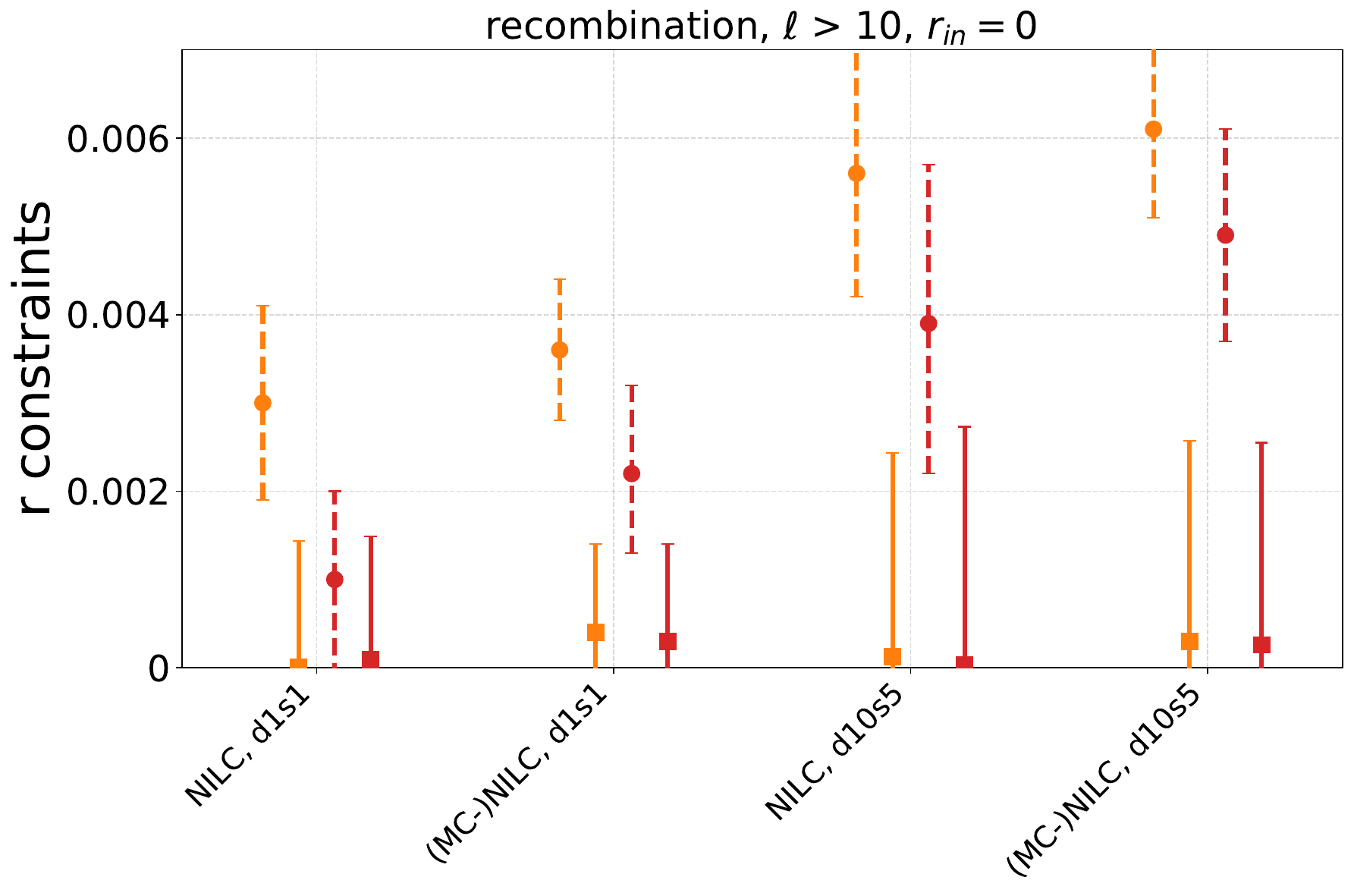} \\
    \includegraphics[width=0.495\textwidth]{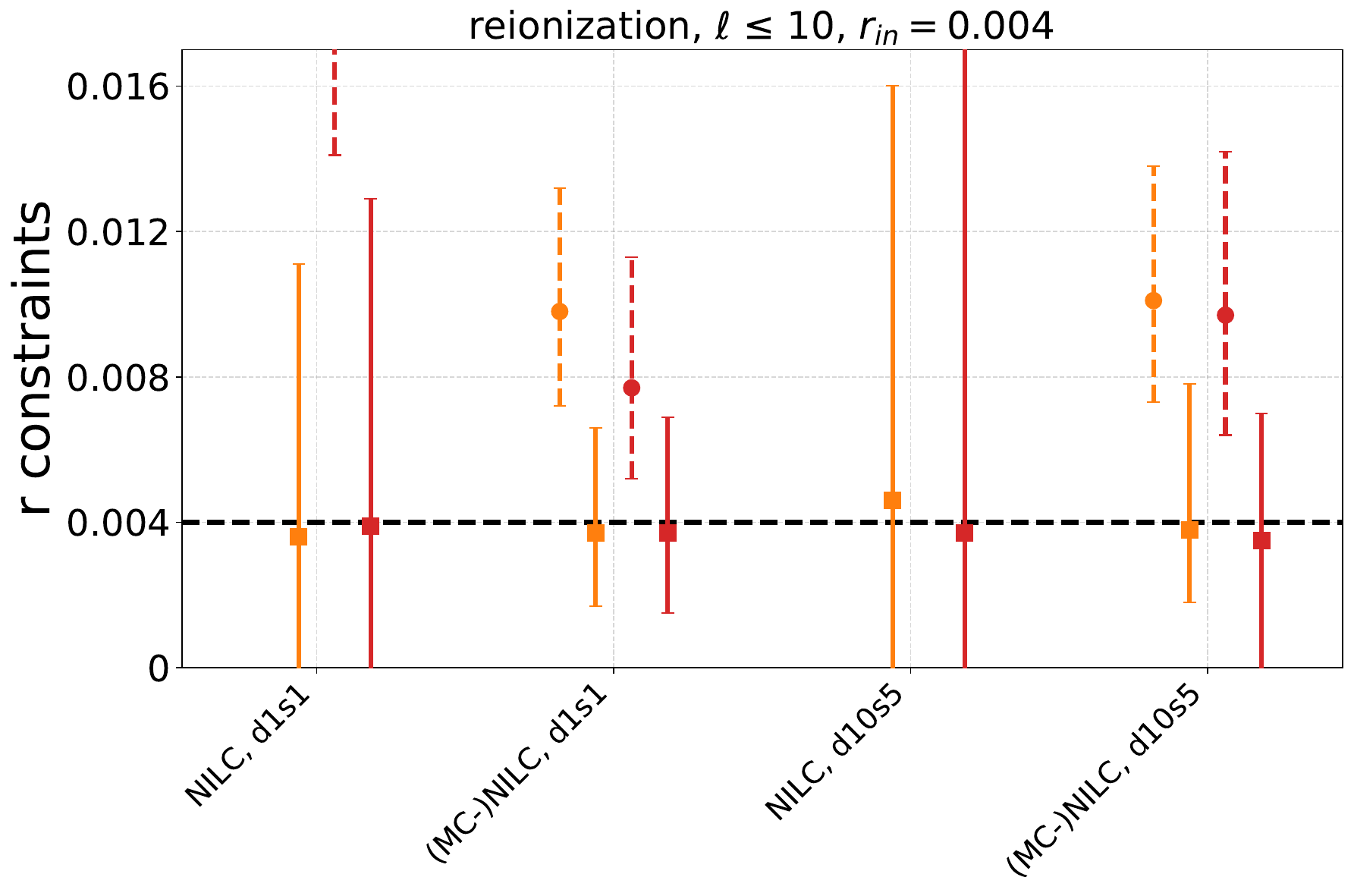}
    \includegraphics[width=0.495\textwidth]{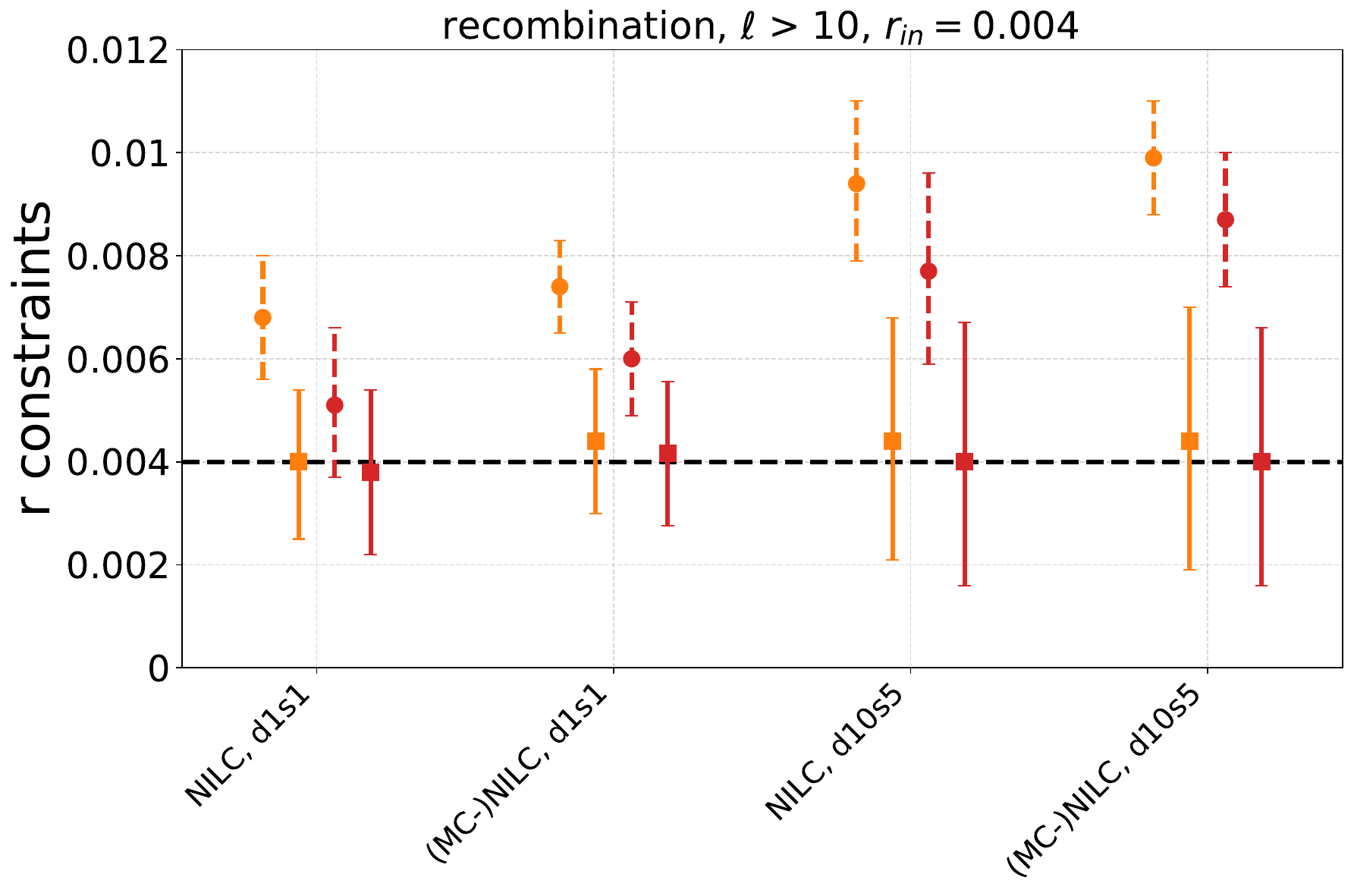} \\
    \includegraphics[width=0.495\textwidth]{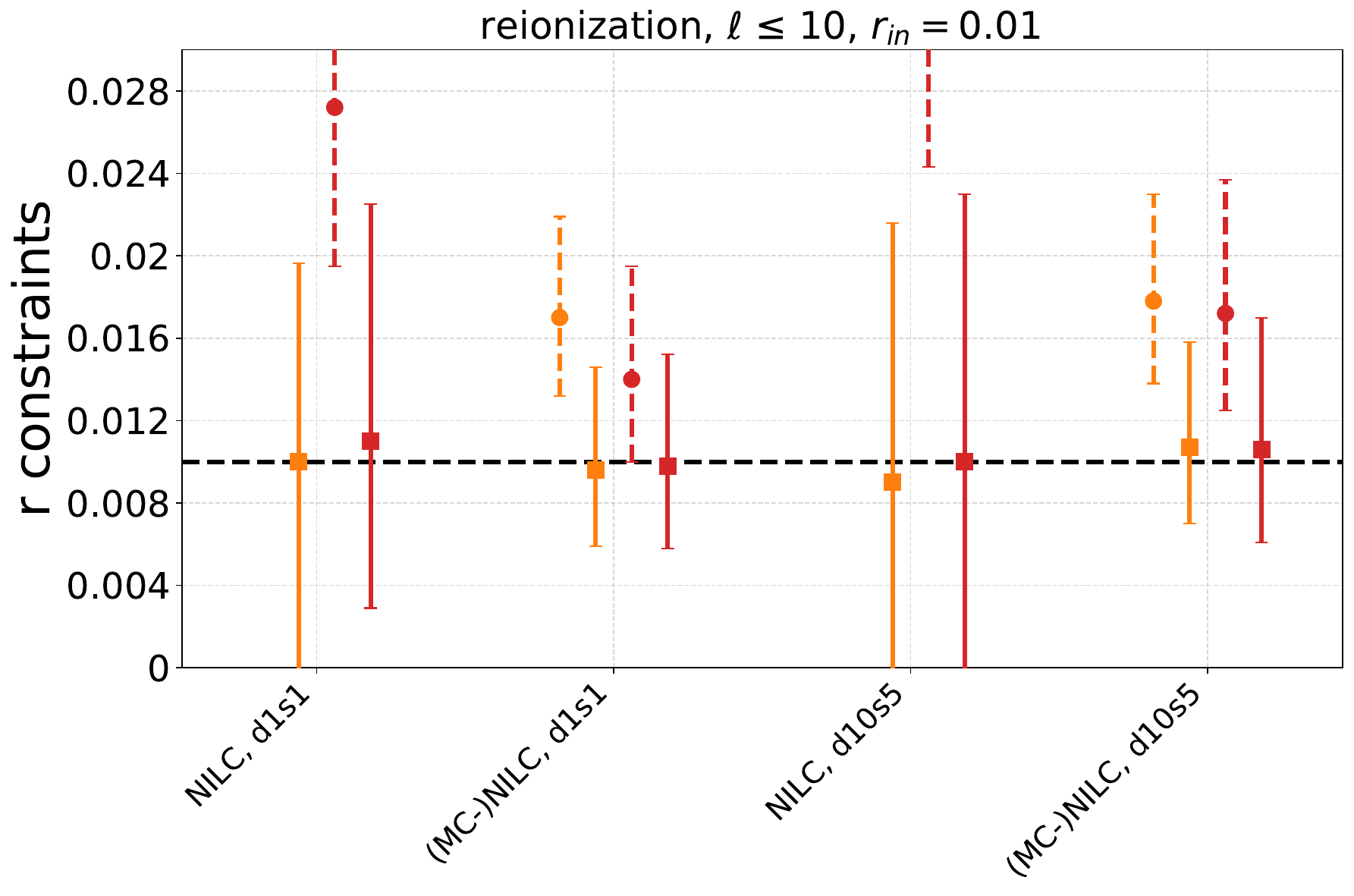}
    \includegraphics[width=0.495\textwidth]{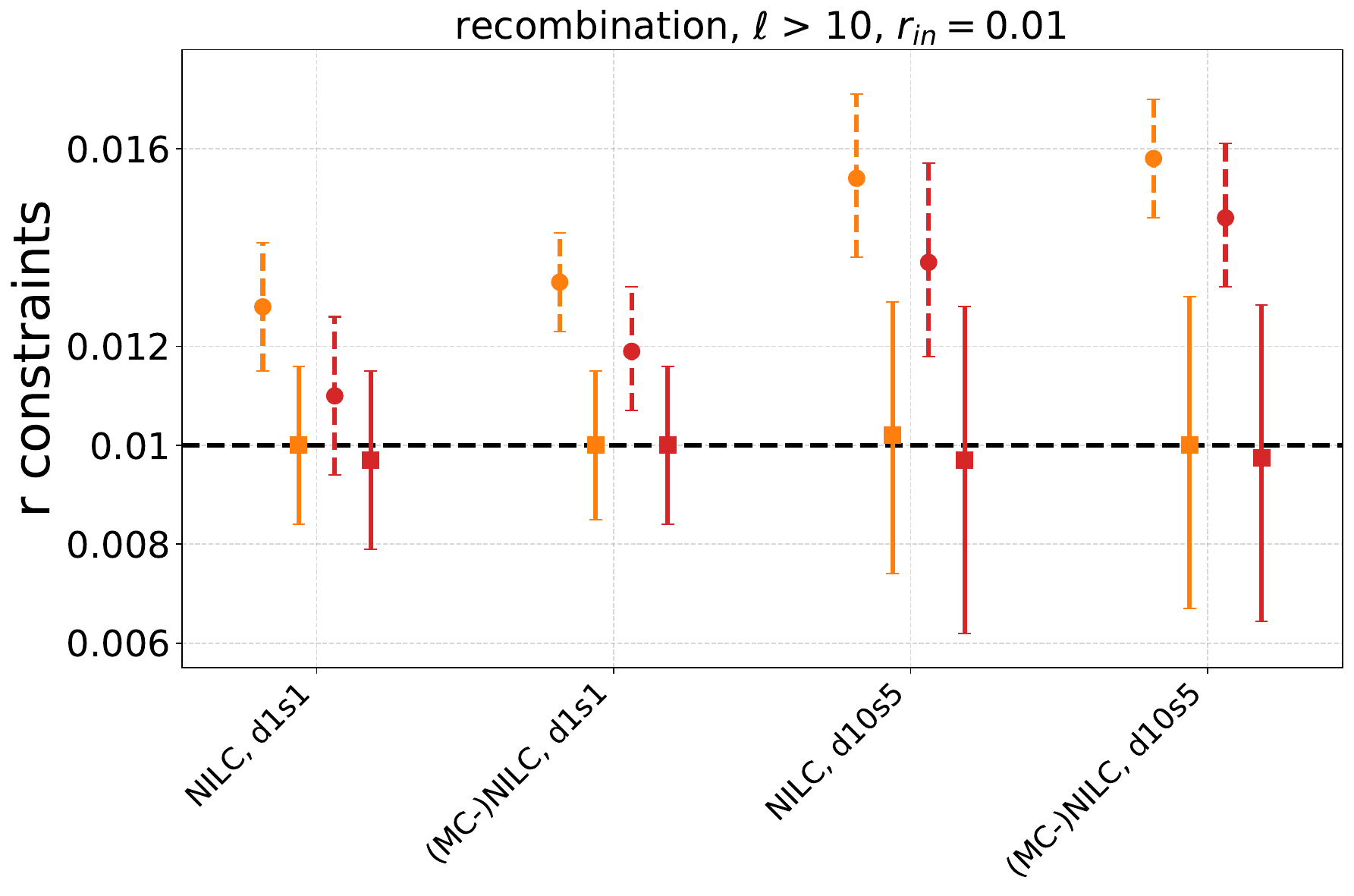}
	\caption{Constraints on the tensor-to-scalar ratio, shown as best-fit values with their corresponding $68\%$ confidence intervals. The left and right panels respectively display results obtained when considering only the reionization ($\ell \le 10$) and recombination ($\ell > 10$) bumps of the tensor spectrum. Results are reported for all combinations of component-separation pipelines (NILC and (MC-)NILC), foreground models (\texttt{d1s1} and \texttt{d10s5}), and masking strategies (\texttt{GAL60} and \texttt{GAL40}). Different rows correspond to the various input tensor-to-scalar ratios $r_{\mathrm{in}}$ used in the simulations: $0$ (top), $0.004$ (middle), and $0.01$ (bottom). Constraints shown with circles and dashed lines correspond to cases without marginalization, whereas those with squares and solid lines correspond to cases in which a template of the foreground residuals is included in the spectral model. Some of the constraints for the NILC cases in the reionization regime are missing, as the corresponding biases are significant and shift the high-probability interval beyond the considered range for the vertical axis.}
	\label{fig:reio_recomb}
\end{figure}

To further evaluate the robustness of the proposed procedure, we repeat the same analysis for all the considered cases, this time separately examining the power at the reionization and recombination peaks of the tensor spectrum. In practice, this is achieved by sampling the likelihood in Equation~\ref{eq:like}, restricting the multipole ranges to $\ell \leq 10$ and $\ell > 10$, respectively. The corresponding results, obtained with and without marginalization, are presented in Figure~\ref{fig:reio_recomb}, separately for the two peaks and for the different values of $r_{\mathrm{in}}$ considered in this work. Overall, we find that applying marginalization yields unbiased estimates of the tensor-to-scalar ratio for both peaks and across all examined foreground models, masking strategies, and input amplitudes of the tensor perturbations. This demonstrates the effectiveness of the proposed marginalization approach in mitigating residual foreground contamination in both the reionization and recombination regimes. We observe that, as expected, the overall uncertainty on the tensor-to-scalar ratio is significantly reduced at reionization in the (MC-)NILC framework, owing to the substantial decrease in reconstruction noise achieved through the introduction of tailored sky patches, as also illustrated in Figure~\ref{fig:spectra_comp}. The difference in error bars between NILC and (MC-)NILC becomes much smaller at recombination, where the same component-separation approach is employed in both pipelines. We also note the presence of a small residual bias, particularly in the reionization regime at large angular scales, likely arising from inaccuracies in the adopted likelihood formalism, which is not fully valid at very low multipoles in the presence of masking and the resulting multipole correlations.

% The results in Fig.~\ref{fig:map_comp} are shown in the full sky, while Fig.~\ref{fig:spectra_comp} highlights spectra .

\section{Conclusions}
\label{sec:concl}
Several previous forecast analyses for future CMB experiments have shown evidence of residual foreground contamination in the reconstructed CMB polarization field, which can bias the inferred cosmological parameters \cite{PTEP,cMILC,MCNILC,2024A&A...686A..16W,PICO,2025Rizzieri,CMBS4_forecast}. As discussed in Section~\ref{sec:intro}, this issue can be addressed in three distinct (though not mutually exclusive) ways:  
(i) by improving component-separation techniques;  
(ii) by refining the masking strategy to better identify and exclude sky regions strongly affected by residuals;  
(iii) by explicitly accounting for such contamination at the power-spectrum level.  

In this work, we mainly focus on the third approach. We note that, in other similar forecast analyses employing alternative map-based component-separation techniques \cite{FGBuster, 2025Rizzieri}, a spectral template of foreground residuals corresponding to the power spectrum of an observed dust-dominated frequency channel has already been successfully incorporated into the cosmological likelihood \cite{PTEP, 2024A&A...686A..16W}.  

In this context, our work represents an advancement in the treatment of foreground residuals by introducing a more agnostic and coherent method for deriving the residual foreground power-spectrum template directly from the data. This template is then incorporated into the cosmological likelihood to mitigate potential biases on cosmological parameters induced by residual foregrounds.  

We focus on the specific case of constraining the tensor-to-scalar ratio from cleaned CMB $B$-mode maps, as this represents the most contamination-prone observable in future analyses of microwave data. The proposed procedure is simple and intuitive, yet proves both effective and robust when validated on realistic simulations of the microwave sky expected for a \emph{LiteBIRD}-like satellite \cite{PTEP}. 

Cleaned templates of foreground emission at the different observed frequency channels are derived from the multifrequency simulated data set using the well-established blind technique GNILC \cite{GNILC_intro, GNILC}. These templates are then combined with the component separation weights, yielding a cleaned estimate of the spatial distribution of foreground residuals. The corresponding power spectrum, properly debiased from residual noise, is subsequently included in the spectral model of the cosmological likelihood to account for contamination from component-separation residuals in the observed power spectrum (see Equation~\ref{eq:cl_templ}). We consider two slightly different pipelines for the component-separation step, NILC and (MC-)NILC, which differ in the implementation of the minimum-variance combination at large angular scales (see Section~\ref{ssec:compsep} for further details).

We demonstrate in this analysis that the proposed approach, when applied to \lb-like $B$-mode simulations, allows us to recover an unbiased posterior of the tensor-to-scalar ratio, regardless of the assumed input value of $r$ (see Figure~\ref{fig:posteriors_marg} and Table \ref{tab:r_posts}). In the same cases, posteriors obtained without including the foreground residual component in the model show significant biases, highlighting both the effectiveness and the necessity of the proposed approach. This conclusion holds independently of the simulated Galactic foreground model or the adopted masking strategy, underscoring the robustness of the results under varying conditions.

The proposed procedure is general and can be readily extended to other map-based component-separation pipelines, whether parametric \cite{2022MNRAS.517.2855D,2023MNRAS.518.3675D,Commander,2025Rizzieri} or blind \cite{SEVEM,cMILC,ocMILC}, or instrumental configurations, such as the Simons Observatory \cite{SO_2019}, whose latest forecasts reported residual biases in the tensor-to-scalar ratio arising from foreground contamination when using the NILC pipeline \cite{2024A&A...686A..16W}. Such bias has been shown to be mitigated through tailored (idealized) masking \cite{NILC_cutsky}; therefore, the proposed procedure could serve as an alternative means to retain a larger sky fraction, potentially reducing the statistical uncertainty.

The marginalization methodology is based on the reconstruction of cleaned templates of the foreground emission at the various frequency channels used for CMB reconstruction. In practice, it relies on the GNILC approach (see Section~\ref{ssec:compsep}). However, other techniques available in the literature may offer further improvements or enhanced robustness, such as the Wavelet Scattering Transform framework \cite{Delouis2022,Mousset2024} and non-local means methods \cite{nonlocal_means}, owing to their ability to exploit higher-order statistics.

The introduced technique promises to be also valuable for other CMB science cases where large-scale foreground contamination, although less significant compared to the CMB signal, may still affect cosmological analyses—for instance, the inference of the optical depth to the last-scattering surface, $\tau$, from the low-$\ell$ $E$-mode power spectrum.

The inclusion of marginalization can therefore be applied in any situation where contamination from foreground residuals becomes evident, for instance as indicated by higher-order statistical assessments of the reconstructed CMB maps, such as those explored in \cite{Ranucci_MFs}.

\acknowledgments
AC acknowledges Nicoletta Krachmalnicoff and Carlo Baccigalupi for their valuable comments throughout the development of this project. In this work we made use of the following additional software/packages: \texttt{numpy}\footnote{\href{https://numpy.org/}{numpy.org/}}, \texttt{astropy}\footnote{\href{https://www.astropy.org/}{astropy.org/}}, \texttt{matplotlib}\footnote{\href{https://matplotlib.org/}{matplotlib.org/}}.
AC acknowledges partial support by the Italian Space Agency (ASI Grants No. 2020-9-HH.0 and 2016-24-H.1-2018), as well as the RadioForegroundsPlus Project HORIZON-CL4-2023-SPACE-01, GA 101135036 and through the Project SPACE-IT-UP by the Italian Space Agency and Ministry of University and Research, Contract Number 2024-5-E.0.
This work has also received support by the European Union’s Horizon 2020 research and innovation programme under
grant agreement no. 101007633 CMB-Inflate.

\bibliographystyle{JHEP}
\bibliography{biblio}

\end{document}